\documentclass[aps,pra,twocolumn,superscriptaddress,nofootinbib]{revtex4-2}
\usepackage{amsmath}
\usepackage{float}
\usepackage{mathtools}
\usepackage{amssymb}
\usepackage{graphicx}
\usepackage{bm}
\usepackage{makecell}
\usepackage[colorlinks=true, citecolor=blue, linkcolor=blue, urlcolor=blue]{hyperref}
\usepackage{color,dsfont,multirow,booktabs}
\usepackage{float}
\usepackage{braket}
\usepackage{tabularx}
\usepackage{lipsum,booktabs}
\usepackage{placeins}
\usepackage{subfiles}
\usepackage{subcaption}
\usepackage{afterpage}
\usepackage[utf8]{inputenc} 
\usepackage[dvipsnames]{xcolor}
\usepackage{ragged2e}
\usepackage{physics}
\usepackage{multirow}

\makeatletter
\renewcommand{\fnum@figure}{\textbf{Figure~\thefigure}}
\makeatother
\usepackage{orcidlink}
\usepackage[T1]{fontenc}
\usepackage[normalem]{ulem}%  to take the underlines out

\usepackage{physics}
\usepackage{mathtools}
\usepackage{hyperref}
\usepackage{dsfont}
\usepackage{enumitem}
\usepackage{mathrsfs}
\usepackage{calrsfs}
\usepackage{comment}

\DeclareMathAlphabet{\pazocal}{OMS}{zplm}{m}{n}

\newcommand{\ml}[1]{\textcolor{purple}{[{\tt ML}: #1]}}

\newcommand{\UGxIFTiA}{Institute of Theoretical Physics and Astrophysics, Faculty of Mathematics, Physics and Informatics, University of Gda\'nsk, ul. Wita Stwosza 57, 80-308 Gda\'nsk, Poland}

\newcommand{\UGxICTQT}{International Centre for Theory of Quantum Technologies (ICTQT), University of Gda\'nsk, prof. Marii Janion 4, 80-309 Gda\'nsk, Poland}

\def\blk{\color{black}}

\begin{document}
\date{\today}

\title{Which Otto Engine Is the Fastest?}

\author{Idriss Hank Nkouatchoua Ngueya\,\orcidlink{0000-0003-4779-2677}}
\email{idriss.nkouatchoua.ngueya@ug.edu.pl}
\affiliation{\UGxICTQT}
\affiliation{\UGxIFTiA}

\author{Marcin {\L}obejko\,  \orcidlink{0000-0002-7159-5502}}
\affiliation{\UGxIFTiA}

\begin{abstract}
A general thermal machine is characterized by distinct time or energy scales: temperature, coupling to the heat baths, internal free dynamics and interactions, and external driving. We address the question: what parameters determine the speed of a given engine, and how can they be used to reliably compare different types of engine? Specifically, we consider four realizations of the Otto engine, each operating with the same Otto efficiency, and aim to characterize and compare their power outputs. The main insight of this paper is that, irrespective of their very different dynamical implementations, the power of each engine can be factorized into a common characteristic work and an implementation-specific characteristic time. The characteristic work depends only on the internal frequencies of the machine and the temperatures, whereas the characteristic time depends on the coupling strength to the baths and on the implementation-specific driving. This observation allows us to compare the power of different engines through their characteristic times and thereby reduce the relevant parameter space to parameters that capture the dynamical aspects of engine performance. Despite different dynamical implementations, this framework reveals simple common features. In all cases, the maximum operating speed is limited by a common timescale given by the sum of the characteristic thermalization times of the hot and cold baths. Moreover, the four engines fall into two distinct asymptotic classes: implementations in which driving and thermalization occur simultaneously exhibit quadratic scaling, whereas those in which work extraction and thermalization alternate exhibit linear scaling. These results expose general dynamical features that determine the operational speed of quantum Otto engines.
\end{abstract}

\maketitle

\section{Introduction}

Quantum Otto engines have progressed from theoretical models to experimentally realizable thermal machines on a variety of quantum platforms.  Otto-type operation has been demonstrated using a single trapped ion \cite{Rossnagel2016}, nuclear spins controlled by nuclear magnetic resonance \cite{Peterson2019,Assis2019}, nitrogen-vacancy centers in diamond \cite{Klatzow2019}, individual atoms coupled to ultracold atomic reservoirs \cite{Bouton2021,Nettersheim2022}, coupled optomechanical resonators \cite{Sheng2022}, and, more recently, superconducting circuits with engineered thermal reservoirs \cite{SuperconductingOtto2026}. Collectively, these experiments demonstrate that the
basic ingredients of a quantum Otto engine--controllable energy gaps, engineered heat exchange, and finite-time work strokes--can now be realized in markedly different physical systems. They also reveal that the same thermodynamic cycle can be implemented through substantially different driving and
thermalization protocols, which motivates a systematic comparison of their operational performance.

Against this rapidly developing experimental backdrop, the quantum Otto engine remains one of the most widely studied quantum heat-engine models, operating as the quantum analogue of the classical Otto cycle \cite{feldmann1996heat, quan2007quantum, he2009performance, henrich2007quantum, agarwal2013quantum, zhang2014quantum, Lobejko2024PRL, Lobejko2025EquivalenceOD}. Previous studies have extensively addressed finite-time performance, including output power, efficiency at maximum power, and power–efficiency trade-offs \cite{Feldmann2000, Abah2012, Wu2014, AbahLutz2018}. These results highlight the central role of dynamical timescales in determining engine performance. Nevertheless, a systematic comparison of how different dynamical implementations of the same thermodynamic cycle affect the operational speed of an engine remains largely unexplored. Although the trade-off between power and efficiency has received considerable attention, comparatively little work has examined the operational speed of quantum Otto engines across different operating regimes or systematically compared how different dynamical implementations affect this speed.

In this paper, we consider four different dynamical models of the Otto engine, i.e., operating with the same Otto efficiency, based on its two discrete versions: one based on a four-stroke cycle and the other on a two-stroke cycle. Having established the same fixed efficiency for each engine, which is, in principle, temperature-independent, the remaining question is how these engines differ in terms of their power output.

In general, we aim to compare the engines and characterize their power over their full ranges of operation, thereby allowing us to address, within specified parameter regimes, the question of ``which Otto engine is the fastest''. However, even after exactly solving the models, answering this question is not straightforward. This difficulty arises from the generally large parameter space and the conceptual differences among the model-specific parameters characterizing the performance of each engine.

To overcome this difficulty, we introduce a common framework based on the characteristic time of an engine, which is discussed in detail in the next section. For all four realizations considered here, the power can be factorized into a common work contribution and a model-dependent characteristic time,
\begin{equation}
P(\bm \beta, \bm \omega, \bm \tau, \bm\Omega)
=\frac{W(\bm \beta, \bm \omega)}{T(\bm \tau, \bm \Omega)}. \nonumber
\end{equation}
The characteristic work $W(\bm \beta, \bm \omega)$ depends only on the temperatures $\bm \beta$ and internal frequencies $\bm \omega$, whereas $T(\bm \tau, \bm \Omega)$ contains the dynamical parameters associated with thermalization $\bm \tau$, and external driving $\bm \Omega$. Consequently, when the thermodynamic parameters are fixed, comparing the power of the engines reduces to comparing their characteristic times: the engine with the shortest characteristic time produces the highest power.

This factorization allows us to separate the thermodynamic and dynamical aspects of engine performance. It also substantially reduces the number of parameters required for a reliable comparison. We derive the characteristic time for each of the four realizations and express it in terms of dimensionless dynamical parameters. For two of the models, the power is determined by a single dimensionless parameter, while two and three parameters are initially required for the remaining two models, respectively. After further optimization, all four realizations can be represented in a one-dimensional parameter space. This representation covers their full ranges of operation and allows us to identify the minimum characteristic time, determine the behavior at small parameter values, and derive the power-law behavior at large parameter values.

The paper is organized as follows. In Sec.~\ref{sec:characteristic-time}, we introduce the characteristic operational time and show how it provides a common measure for comparing the speed of different Otto engines. In Sec.~\ref{section: discrete models}, we review the discrete four-stroke and two-stroke Otto cycles that serve as reference models for the dynamical implementations considered in this work. In Sec.~\ref{section: Dynamical models}, we introduce the four dynamical realizations and derive their corresponding characteristic times. In Sec.~\ref{section 3}, we analyze and optimize these characteristic times and compare their behavior across the different implementations. Finally, in Sec.~\ref{sec:summary}, we summarize our main results and discuss their implications and possible extensions.

\section{Characteristic time}\label{sec:characteristic-time}
Every quantum heat engine can generally be characterized by four types of time or energy scales: inverse temperatures $\bm \beta$, internal frequencies $\bm \omega$, characteristic equilibration timescales (or, conversely, coupling strengths) $\bm \tau$, and external driving frequencies $\bm \Omega$.

The Otto engine is distinguished by the fact that it operates with an efficiency that depends solely on the internal frequencies of the machine, i.e., $\eta = \eta(\bm \omega)$. In particular, for the most common engines, including those discussed in this paper (see the generalization in Ref.~\cite{Lobejko2025EquivalenceOD}), the two characteristic frequencies, i.e., $\bm \omega = \{ \omega_h, \omega_c \}$, determine the efficiency through the well-known Otto formula:
\begin{equation}
\eta(\bm \omega) = 1 - \frac{\omega_c}{\omega_h},
\end{equation}
with $\omega_h > \omega_c$. 

By contrast, the power of the engine depends, in general, on all the characteristic scales introduced above, such that $P = P(\bm \beta, \bm \omega, \bm \tau, \bm \Omega)$. The main observation concerning the four dynamical realizations of the Otto engine discussed here is as follows. Despite significant differences between their dynamical implementations, the power of the engine can be factorized as
\begin{equation} \label{power_factorization}
P(\bm \beta, \bm \omega, \bm \tau, \bm \Omega) = \frac{W(\bm \beta, \bm \omega)}{T^*(\bm \tau, \bm \Omega)}.
\end{equation}
This factorization separates the thermodynamic and dynamical aspects of the engine performance. The characteristic work
\begin{equation}\label{characteristic_work}
W(\bm \beta, \bm \omega) = \frac{(\omega_h - \omega_c)(e^{-\beta_h \omega_h}-e^{-\beta_c \omega_c})}{\left(1+e^{-\beta_h \omega_h} \right) \left(1+e^{-\beta_c \omega_c} \right)}
\end{equation}
depends only on the temperatures and the energy gaps of the working medium and is therefore determined entirely by static thermodynamic quantities. The characteristic operational time $T^*(\bm \tau, \bm \Omega)$, on the other hand, is specific to each dynamical implementation and contains only dynamical quantities: the equilibration timescales associated with the coupling to the thermal reservoirs and the implementation-specific interaction or driving timescales. The direction of operation is determined by the sign of the characteristic work  $W(\bm \beta,\bm \omega)$. The machine operates in the heat-engine regime with positive work production $W>0$, when
\begin{equation}
\omega_h>\omega_c \quad \& \quad \beta_h\omega_h<\beta_c\omega_c.
\end{equation}
As shown in the next section (Section~\ref{section: discrete models}), the quantity $W(\bm \beta, \bm \omega)$ is precisely the work produced in a single cycle by the discrete four-stroke and two-stroke versions of the Otto engine. Therefore, in what follows, we set $W(\bm \beta, \bm \omega) \equiv W_{\rm discrete}$. Moreover, in the two-stroke implementation, $W(\bm \beta, \bm \omega)$ is also the ergotropy of the active product of two-level Gibbs states (when $\beta_h \omega_h < \beta_c \omega_c$ and $\omega_h > \omega_c$).

The characteristic time $T^*(\bm \tau, \bm \Omega)$ is therefore the main object of interest in this paper. According to Eq.~\eqref{power_factorization}, we can compare the ``speed'' of the engines solely in terms of $T^*(\bm \tau, \bm \Omega)$, thereby reducing the number of relevant parameters to those describing the dynamical part: the coupling to the external baths, characterized by the universal equilibration timescales $\bm \tau$, and the characteristic interaction or driving frequencies $\bm \Omega$, which are specific to each implementation.

\section{Discrete models}\label{section: discrete models}
We start with an overview of two models of the discrete Otto engine, namely the four-stroke and two-stroke ones. These models serve as reference points for the main subject of this paper, focusing on dynamical realization of these engines. In particular, we show that both discrete models produce the same work per fully thermalized cycle, and that this work is precisely the characteristic work $W(\bm{\beta},\bm{\omega})$ defined in Eq.~\eqref{characteristic_work}. We therefore identify it with $W_{\rm discrete}$. In the following section, we then construct two dynamical realizations associated with each of these discrete engines.

\subsection{Four-stroke Otto cycle (single qubit)}\label{section: Discrete fours-stroke engine}
The first discrete version of the Otto engine consists of a single two-level system with energy levels $\ket 0$ and $\ket 1$, the so-called `qubit', interacting with two baths at different inverse temperatures $\beta_h$ and $\beta_c$ ($\beta_h < \beta_c$). This is the standard textbook description of the Otto cycle, which is composed of four strokes: 
\begin{enumerate}[label=(\roman*)]
        \item Hot bath thermalization for a fixed energy gap $\omega_h$;
        \item Isolated level transformation: $\omega_h \to \omega_c$;
        \item Cold bath thermalization for a fixed energy gap $\omega_c$;
        \item Isolated level transformation: $\omega_c \to \omega_h$;
\end{enumerate}
Assuming full thermalization to the Gibbs states, the total heat exchanged with the hot and cold baths can be defined as
\begin{equation}
    Q_h = \Tr[{H}_h (\rho_h - \rho_c)], \quad Q_c = \Tr[{H}_c (\rho_h - \rho_c)],
\end{equation}
where ${H}_k = \omega_k \dyad{1}$ and $\rho^{\rm eq}_k = e^{-\beta_k H_k}/\Tr[e^{-\beta_k H_k}]$ (for $k=h,c$). One can show that
\begin{equation}\label{eq:heats}
    Q_h = \omega_h \delta p, \quad Q_c = - \omega_c \delta p  
\end{equation}
where 
\begin{equation}\label{eq:deltap}
    \delta p = \frac{e^{-\beta_h \omega_h}-e^{-\beta_c \omega_c}}{\left(1+e^{-\beta_h \omega_h} \right) \left(1+e^{-\beta_c \omega_c} \right)}.
\end{equation}
The work performed by the engine in the cycle follows from the first law:
\begin{equation}\label{eq: discrete work}
    W_{\rm discrete} = Q_h + Q_c = (\omega_h - \omega_c) \delta p.
\end{equation}
The corresponding efficiency of the engine after one cycle is therefore
\begin{equation}\label{eq:eff}
\eta=\frac{W}{Q_h}=1-\frac{\omega_c}{\omega_h}.
\end{equation}
\blk
\subsection{Two-stroke Otto cycle (two qubits)}
The second discrete version of the Otto engine is, in contrast, realized in just two strokes, but at the cost of using two qubits simultaneously instead of one. The model has been extensively studied in \cite{Lobejko2024PRL, PhysRevE.110.044120, Lobejko2025EquivalenceOD}. Here, the initial state is a tensor product of two Gibbs states, i.e., $\rho_h \otimes \rho_c$, and the strokes are as follows:
\begin{enumerate}[label=(\roman*)]
\item \textit{Work stroke}: The qubits are decoupled from the baths and are externally driven, resulting in a swap process, namely: 
\begin{gather}
    \rho_h \otimes \rho_c \xrightarrow{\rm work\text{-}stroke} {S} \rho_h \otimes \rho_c {S}^\dag, 
\end{gather}
where 
\begin{equation}
    {S} = \dyad{10}{01} + \dyad{01}{10} + \dyad{00}{00} + \dyad{11}{11}.
\end{equation}
As the process is unitary, this results in work extraction:
\begin{equation}
    W_{\rm discrete} = \Tr[({H}_h + {H}_c) (\rho_h \otimes \rho_c - {S} \rho_h \otimes \rho_c {S}^\dag)].
\end{equation}
\item \textit{Heat stroke}: The qubits are reconnected to their respective baths, and they equilibrate to the original Gibbs states, i.e., 
\begin{equation}
   {S} \rho_h \otimes \rho_c {S}^\dag \xrightarrow{\rm heat\text{-}stroke}  \rho_h \otimes \rho_c,
\end{equation}
which induces the corresponding heat flows, given by:
\begin{equation}
    Q_{h,c} = \Tr[H_{h,c} (\rho_h \otimes \rho_c - {S} \rho_h \otimes \rho_c {S}^\dag)].
\end{equation}
\end{enumerate}
As can be shown, the expressions for heat, work and efficiency are identical to those derived for the four-stroke version, i.e., Eqs.~\eqref{eq:heats}, \eqref{eq: discrete work}, and \eqref{eq:eff}, respectively. 

\section{Dynamical models}\label{section: Dynamical models}
We now turn to the main subject of this paper, namely the dynamical implementations of the above discrete models. In the following, we introduce four different dynamical models of the Otto engine. Two of them are implemented with a working medium consisting of a single qubit and are realized via energy level transformations, which we refer to as \textit{longitudinal coupling}. The remaining two models involve two qubits and are based on the swap process, called \textit{transverse coupling}. 

In all of the models presented, we use the standard Markovian dynamics to describe the coupling of the engine to hot and cold thermal environments ($k = h,c$), where the dissipative part takes the GKLS form:
\begin{align} \label{dissipator}
\pazocal{L}_k \rho = \Gamma_k^+ \pazocal{D}[\sigma_+]\rho + \Gamma_k^- \pazocal{D}[\sigma_-]\rho,% \pazocal{D}\left[\rho(t)\right],
\end{align}
where 
\begin{equation}
\pazocal{D}[A]\rho = A \rho A^\dag - \frac{1}{2} \{A^\dag A, \rho \}.
\end{equation}
and $\sigma_+ = \dyad{1}{0}$, $\sigma_- = \left(\sigma_+\right)^\dag$. 

In particular, for a qubit weakly coupled to a thermal bath within the Davies framework \cite{BreuerPetruccione2002, Davies1974}, the transition rates are given by:
\begin{equation} \label{coupling_rates}
\Gamma_\pm = \gamma^\pm_k(\omega) = \int dt e^{\pm i\omega t} \langle B_k(t) B_k(0)\rangle,
\end{equation}
and satisfy the detailed balance condition: $\gamma_k^+ (\omega) / \gamma_k^- (\omega) = e^{-\beta_k \omega}$. Based on that, we introduce a characteristic equilibration time given by:
\begin{align}\label{eq: thermalization timescales}
\tau_k = \frac{1}{\gamma_k^+ (\omega_k) + \gamma_k^{-}(\omega_k)}. 
\end{align}
Also, we introduce the asymmetry parameter: 
\begin{equation}\label{kappa}
    \kappa=\frac{2\sqrt{\tau_h \tau_c}}{\tau_h+\tau_c} \in (0,1],
\end{equation}
which quantifies the imbalance between the timescales of equilibration of the hot and cold reservoirs, where $\kappa = 1$ is for the symmetric case, $\tau_h = \tau_c$.

\subsection{Longitudinal coupling}\label{Longitudinal coupling}

\begin{figure*}[!t]

\centering

    % ---------------- MAIN FIGURE ----------------
 %\includegraphics[width=0.8\textwidth]{Presentation11.pdf}
\includegraphics[ width=0.7\textwidth, trim=0 150 0 0, clip]{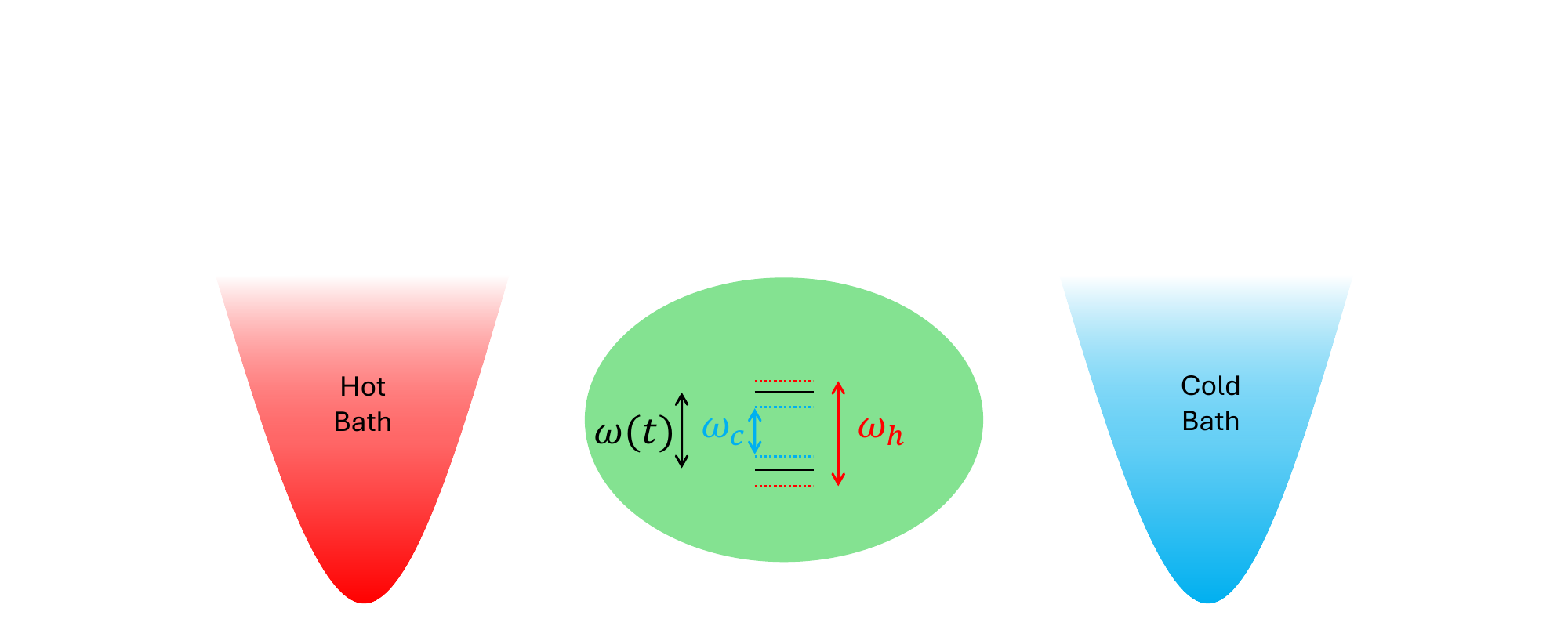}
\begin{center}
\small Longitudinal coupling
\end{center}

% ---------------- SUBFIGURES ----------------
\begin{subfigure}{0.4\textwidth}
        \centering
        \includegraphics[width=\linewidth]{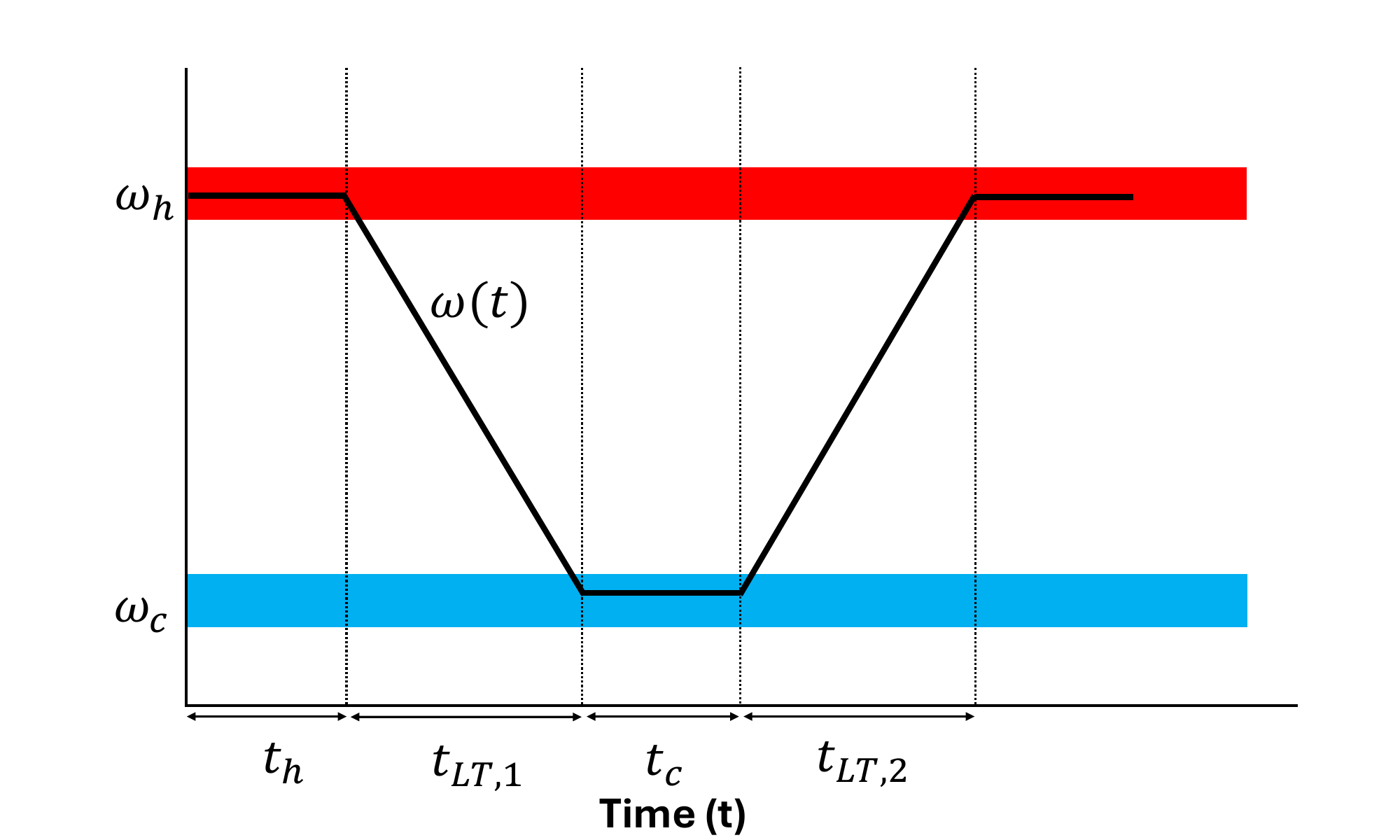}
        \caption{Four-stroke model}
        \label{fig:sub1}
\end{subfigure}
    %\hfill
\begin{subfigure}{0.4\textwidth}
        \centering
        \includegraphics[width=\linewidth]{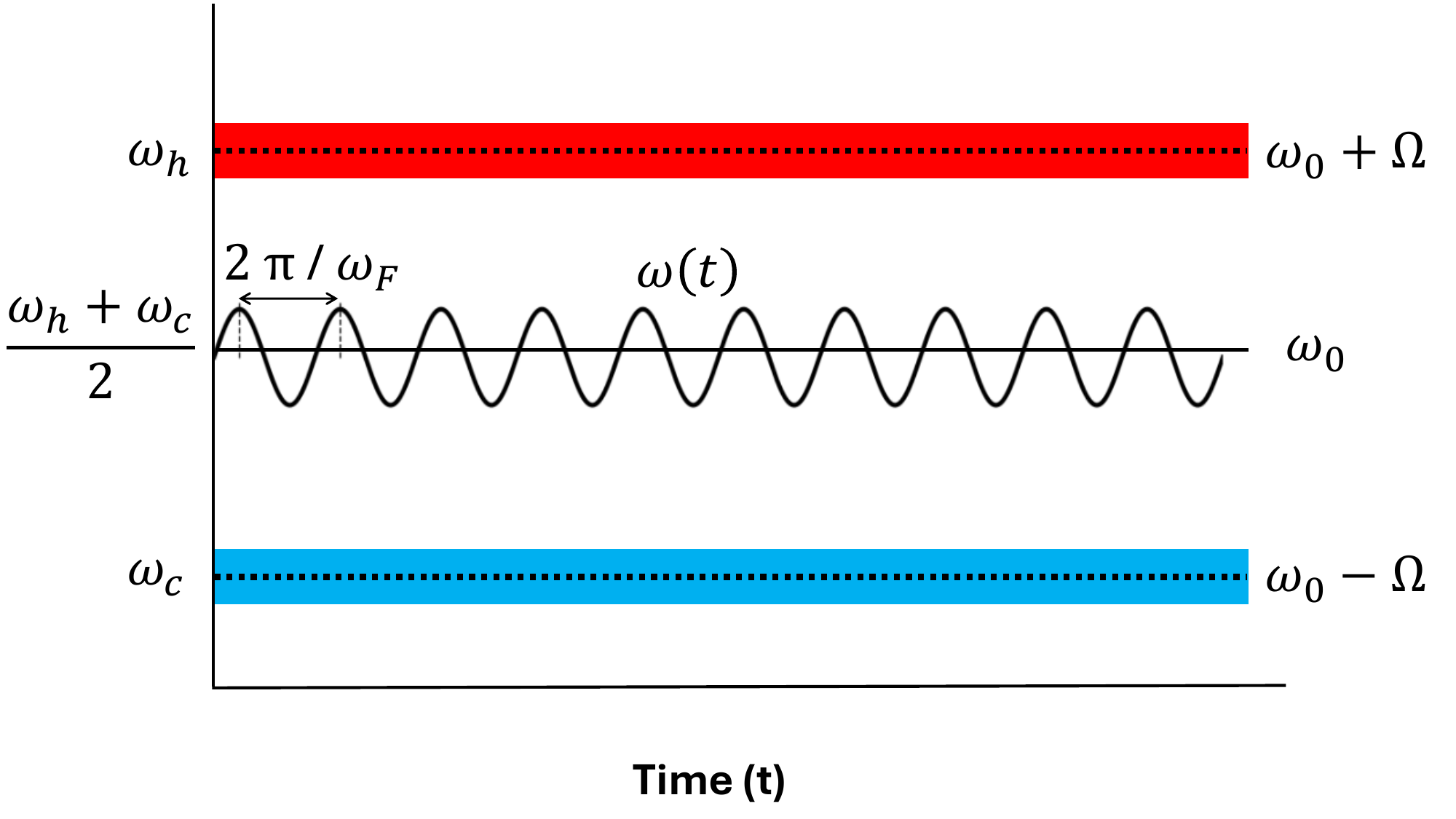} 
        \caption{Floquet-inspired model}
        \label{fig:sub2}
\end{subfigure}

    %\vspace{0.3cm}

\caption{ The figures illustrate a quantum Otto engine based on a single qubit serving as the working medium and subject to an external time-dependent drive. Panel (a) depicts the conventional four-stroke realization, where the qubit frequency is modulated between two values through a sequence of expansion and compression strokes. In this case, the frequency varies linearly within each stroke. Panel (b) presents a Floquet-inspired realization in which the qubit frequency undergoes a continuous periodic modulation, with a driving period shorter than the characteristic relaxation timescale of the system.}

\label{fig:fullpage}

\end{figure*}

We start first with two models which focus on a single qubit influenced by an external longitudinal modulation. In this case, the Hamiltonian reads 
\begin{align} \label{logitudinal_modulation}
{H}_{\rm{long}}(t)=\omega(t) \ket{1}\bra{1},
\end{align}
where $\omega(t)$ represents the external modulation applied to the qubit system. 

\subsubsection{Dynamical four-stroke model}\label{Dynamical four-stroke model}
Our first dynamical model is a finite-time implementation of the discrete four-stroke Otto cycle introduced in Sec.~\ref{section: Discrete fours-stroke engine}. That is, the working medium (qubit) is successively coupled to each bath, with a level transformation in between ,with isolated level-transformation strokes between the two thermalization strokes. In this dynamical setting, both thermalization and level transformation occur over a finite time.

We make the following assumptions. First, the longitudinal modulation $\omega(t)$ (Eq. \eqref{logitudinal_modulation}) is periodic, such that $\omega(t)=\omega(t+\tau_{\rm 4-stroke})$, where $\tau_{\rm 4-stroke}$ is the period of a single cycle. Second, the modulation is slow compared to the thermalization timescale, such that the dominant frequencies in the Fourier analysis of $\omega(t)$ are slower than $1/\tau_{h,c}$. %, i.e., $\dot \omega(t) \ll 1/\tau_{h,c}^2$. 
This condition allows us to assume that the dissipator in Eq. \eqref{dissipator} is defined for the instantaneous frequency $\omega(t)$, namely 
\begin{equation} 
\pazocal{L}_{k}[\omega(t)]\rho = \gamma_k^+(\omega(t)) \pazocal{D}[\sigma_+]\rho + \gamma_k^-(\omega(t)) \pazocal{D}[\sigma_-]\rho . \end{equation} 
Thus, the full dynamics is governed by the following master equation: \begin{align} 
\dot{\rho}(t) =-i \left[{H}_{\rm long}(t), \rho(t) \right] + \sum_{k=h,c} \pazocal{L}_{k}[\omega(t)]\rho(t). 
\end{align} Third, as depicted in Fig. \ref{fig:sub1}, in the realization of the four strokes, we assume non-overlapping spectral densities of the hot and cold baths, centered at the frequencies $\omega_h$ and $\omega_c$, respectively, with narrow bandwidths (i.e., $\Delta_{h,c} \ll \omega_{h,c}$). Then, during modulation of $\omega(t)$, the system undergoes a sequence of separated thermalizations at constant frequency and level transformations. Specifically, the four-stroke process is divided into four stages: \begin{enumerate}[label=(\roman*)] \item Hot-bath thermalization: $\omega(t) = \omega_h$ with duration $t_h$; \item Isolated level transformation: $\omega(t): \omega_h \to \omega_c$ with duration $t_{\rm LT, 1}$; \item Cold-bath thermalization: $\omega(t) = \omega_c$ with duration $t_c$; \item Isolated level transformation: $\omega(t): \omega_c \to \omega_h$ with duration $t_{\rm LT, 2}$. \end{enumerate} Here, $\tau_{\rm 4-stroke} = t_h + t_c + t_{\rm LT}$ and $t_{\rm LT} = t_{\rm LT, 1} + t_{\rm LT, 2}$. Since we assume a diagonal form of the density matrix $\rho(t)$, the state does not change during the level transformations. Moreover, as long as the process is relatively slow, the profile can be arbitrary; in Fig. \ref{fig:sub1}, we show the simplest linear profile. 

By contrast, during thermalizations, the state follows purely dissipative dynamics, i.e., $\dot \rho(t) = \pazocal{L}_{h,c} [\omega(t)] \rho(t)$. Finally, due to the narrow bandwidths of the spectral densities, we may also assume that during thermalizations $\omega(t) \approx \omega_{h,c}$, and the state evolves as follows: 
\begin{align}
\dot \rho(t) = \pazocal{L}_{k}(\omega_k)\rho(t) = \Big(\rho^{\rm eq}_{k}-\rho(t)\Big)/\tau_{k}.
\end{align}

According to this, in the $m$-th cycle the state transforms as:
\begin{align}\label{On-off sequence}
    \rho_0^{(m)}
&\xrightarrow{t_h}
\rho_1^{(m)} =  \Lambda^h_{t_{h}}[\rho_0^{(m)}]
\xrightarrow{t_{\rm LT,1}} 
\rho_1^{(m)} \\ 
&\xrightarrow{t_{c}} \rho_2^{(m)} = \Lambda^c_{t_{c}} [\rho_1^{(m)}] \xrightarrow{t_{\rm LT,2}} \rho_2^{(m)}
\end{align}
where $\Lambda^{k}_{t} = e^{\pazocal{L}_{k}(\omega_k) t}$.
We evaluate the heat transfers for the $m$th cycle of the process, obtaining the following formulas:  
\begin{gather}
    W_{\rm 4-stroke}^{(m)} = (\omega_h - \omega_c) \delta p^{(m)}   \\
    {Q^{h \ (m)}_{\rm 4-stroke}} = \omega_{h} \delta p^{(m)}, \quad {Q^{c \ (m)}_{\rm 4-stroke}} = -\omega_{c} \delta p^{(m)}, 
\end{gather}
where $\delta p^{(m)} = \bra{1} \rho_1^{(m)} -\rho_0^{(m)} \ket{1}$ is the change in the excited state population of the qubit (see Appendix \ref{model: 4-stroke engine}). In the periodic steady state (limit cycle) regime, obtained as the number of cycles tends to infinity, we obtain the following expression for the engine power:
\begin{align}
    &P_{\rm 4-stroke} = \frac{1}{\tau_{\rm 4-stroke}}  \lim_{m \to \infty} W^{(m)}_{\rm 4-stroke} \\ &= \frac{\left(1 - e^{ -t_h/\tau_h} \right) \left( 1 - e^{ -t_c/\tau_c} \right) }{(t_h+t_c+t_{\rm LT})(1 - e^{-({ t_h/\tau_h} +  t_c/\tau_c)})} W_{\rm discrete} \label{eq:power_4stroke}
\end{align}
where $W_{\rm discrete}$ is given by Eq. \eqref{eq: discrete work}. Similarly, the efficiency of the engine is given by the Otto formula:
\begin{equation}
    \eta = \lim_{m \to \infty} \frac{W_{\rm 4-stroke}^{(m)}}{Q^{h \ (m)}_{\rm 4-stroke}} = 1 - \frac{\omega_c}{\omega_h}.
\end{equation}

From Eq. \eqref{eq:power_4stroke}, we extract the characteristic time $T^*_{\rm 4-stroke}$ of the dynamical version of the four-stroke Otto engine, namely, the time that our engine operating at the stationary power \(P_{\rm 4-stroke}\) would require to produce the reference work $W_{\rm discrete}$:
\begin{equation}\label{eq:four-stroke time}
  T^*_{\rm 4-stroke} = \frac{(t_h+t_c+t_{\rm LT})(1 - e^{-({ t_h/\tau_h} +  t_c/\tau_c)})}{\left(1 - e^{ -t_h/\tau_h} \right) \left( 1 - e^{ -t_c/\tau_c} \right) }
\end{equation}
The finite-time population dynamics underlying Eq.~\eqref{eq:power_4stroke} has previously been studied in the optimization of four-stroke Otto engines governed by master equations \cite{Feldmann1996,Dann2020}. In particular, the work reduction due to incomplete thermalization obtained in these works is equivalent to the factor appearing in Eq.~\eqref{eq:power_4stroke}. Indeed, their polarization-based reference work, $W_{\rm Otto}=\Delta \Omega \Delta \bar{S}$, coincides with $W_{\rm discrete}$ upon identifying $\omega_k=\hbar \Omega_k$ and $\Delta \bar{S}=\hbar \delta p$, while their relaxation rates correspond to $\Gamma_k=\tau_k^{-1}$. Consequently, maximizing the power of this four-stroke engine is equivalent to minimizing the characteristic operational time in Eq.~\eqref{eq:four-stroke time}. Here, we recast this optimization in terms of a work-normalized operational timescale in order to place the four-stroke engine on the same footing as the other dynamical implementations considered below.
\\

The characteristic time depends on several timescales associated with thermalization and level transformations. For convenience, we recast this expression in terms of the following dimensionless parameters:
\begin{align}
s_{\rm LT}=\frac{t_{\rm LT}}{\sqrt{\tau_h \tau_c}},\quad s_h=\frac{t_h}{\sqrt{\tau_h \tau_c}},\quad s_c=\frac{t_c}{\sqrt{\tau_h \tau_c}},
\end{align}
such that
\begin{align}\label{eq:recast-four-stroke time}
\frac{T^*_{\rm 4-stroke}}{\tau_h + \tau_c}  =\frac{\kappa}{2} \frac{(s_{\rm LT} + s_h + s_c) \left(1 - e^{ -(\sigma s_h + s_c/\sigma)} \right)}{\left(1 - e^{ -\sigma s_h} \right)\left(1 - e^{ -s_c/\sigma} \right)}
\end{align}
where 
\begin{equation}\label{kappa_and_sigma}
    \sigma = \sqrt{\frac{\tau_c}{\tau_h}}, \quad \kappa = 2\left(\sigma + \frac{1}{\sigma} \right)^{-1}.
\end{equation}
Here, $\kappa$ is equal to the definition provided in Eq. \eqref{kappa}.

\begin{figure*}[!t]
\centering
\includegraphics[ width=0.7\textwidth, trim=0 150 0 0, clip]{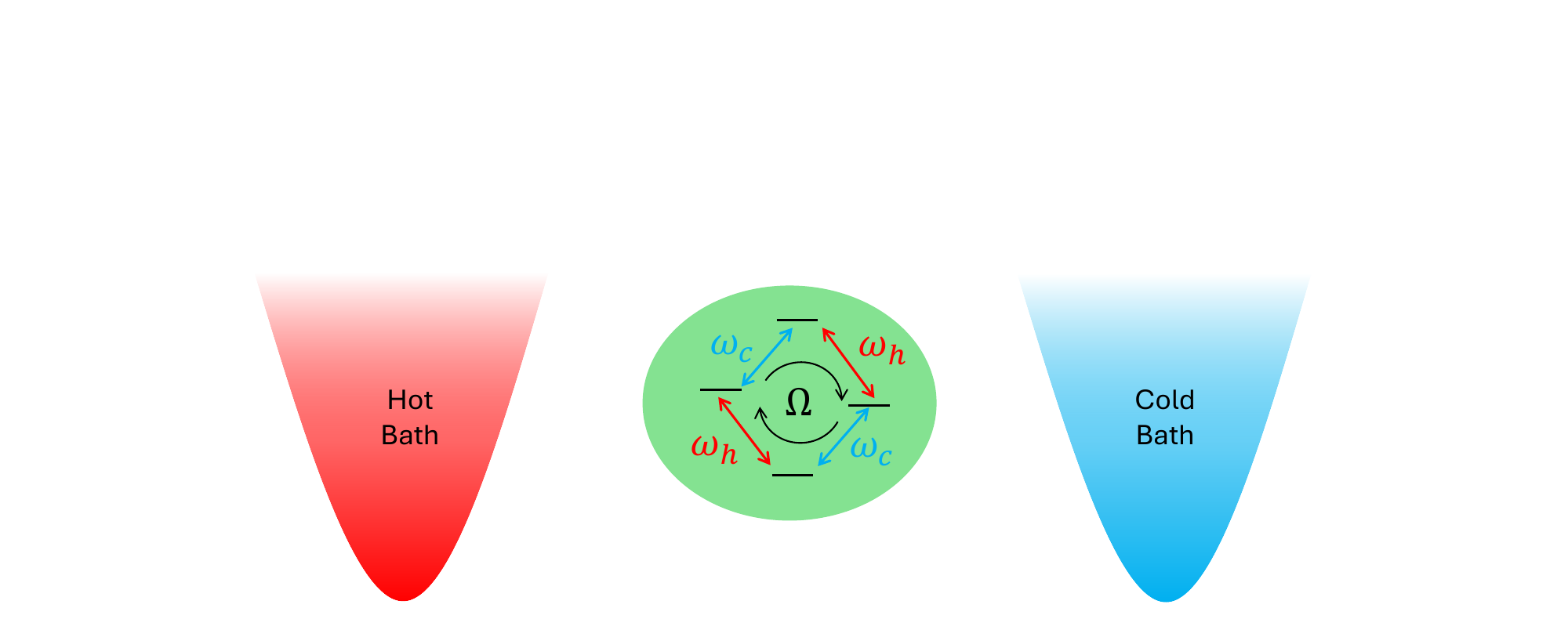}
\begin{center}
\small Transverse coupling
\end{center}
\begin{subfigure}{0.4\textwidth}
\centering
\includegraphics[width=\linewidth]{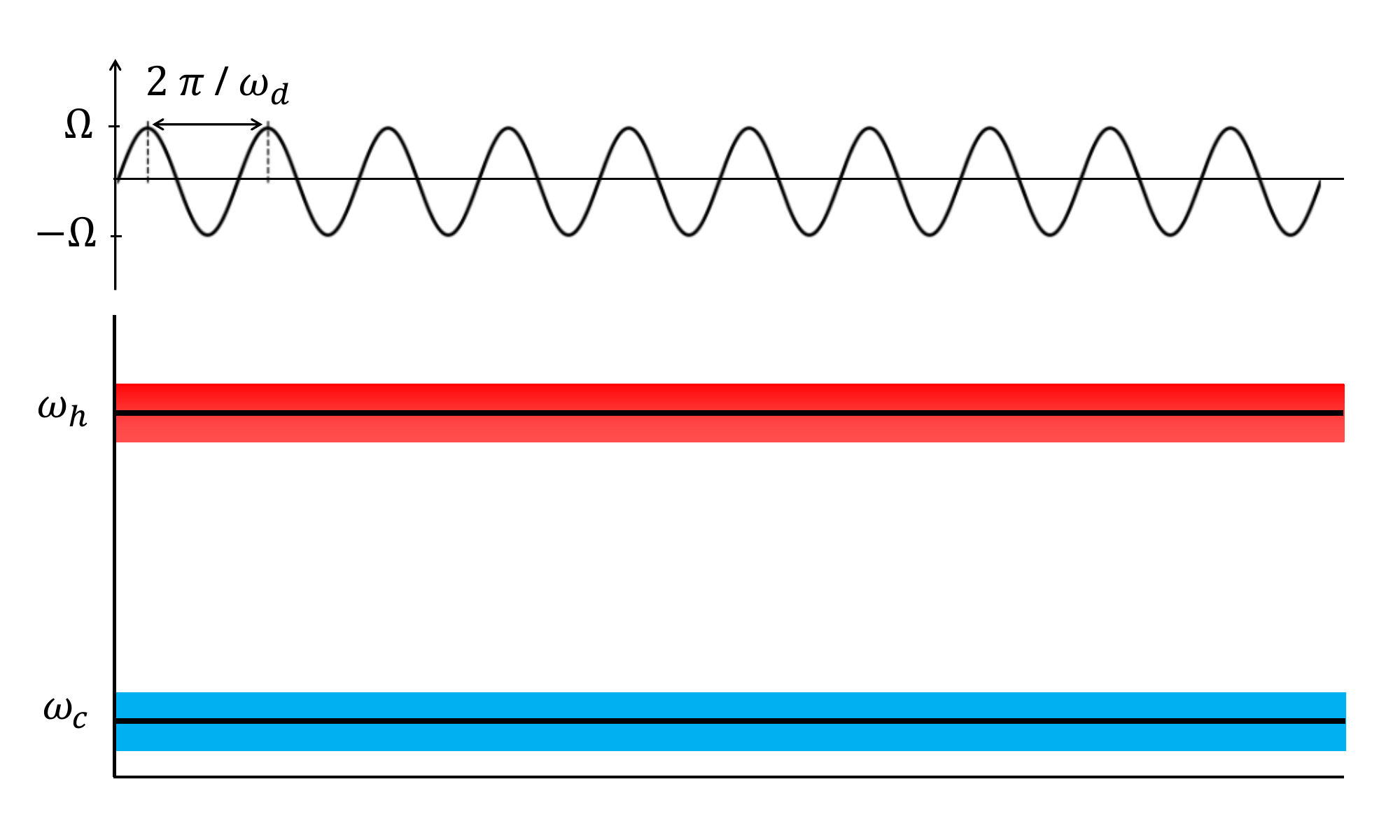}
\caption{Always-on model}
\label{fig:sub5}
\end{subfigure}
%\hfill
\begin{subfigure}{0.4\textwidth}
\centering
\includegraphics[width=\linewidth]{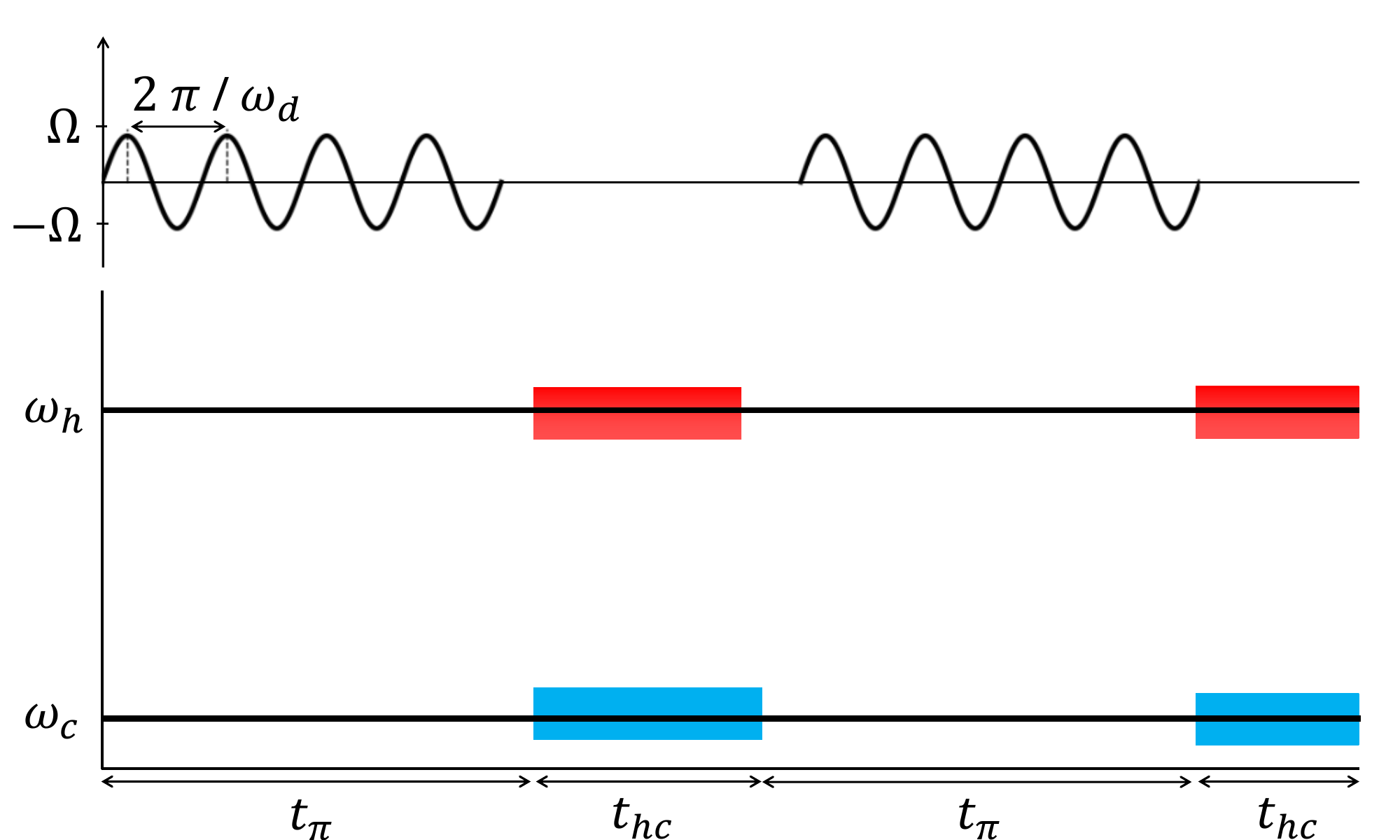}
\caption{On-off model}
\label{fig:sub6}
\end{subfigure}
\caption{The figures above illustrate a quantum Otto engine based on two interacting qubits serving as the working medium and coupled to independent hot and cold thermal reservoirs. The qubits interact through an externally controlled transverse coupling. Panel (a) depicts the always-on realization, where the coupling remains active throughout the operation and the qubit frequencies undergo continuous periodic modulation. Panel (b) presents the on-off realization, in which the interaction is periodically switched on and off, generating a sequence of discrete work strokes separated by thermalization stages.}
\label{fig:fullpage}
\end{figure*}

\subsubsection{Floquet-inspired model} \label{section: Floquet model}

Let us now consider the second model, which is based on the same longitudinal coupling. In contrast to the previous model, here we consider a fast modulation $\omega(t)$ relative to the thermalization timescales We consider a periodic modulation $\omega(t)$ whose frequency is much larger than the dissipative relaxation rates. In particular, for simplicity, we consider the harmonic profile
\begin{equation}
\omega(t) = \omega_0 + \Omega \sin(\omega_{\rm F} t),
\end{equation}
where $\omega_0$ represents the bare frequency of the qubit, $g$ is the modulation strength, and $\omega_{\rm F}$ is the modulation frequency. Then, in the regime $\omega_{\rm F} \gg 1/\tau_{h,c}$, the system can be described using Floquet theory \cite{ASENS_1883_2_12__47_0, PhysRev.138.B979, PhysRevA.7.2203}. In particular, this model was studied in \cite{PhysRevE.87.012140} to design a universal heat engine.

In this model, the fast modulation of $\omega(t)$ introduces harmonics of the driving frequency that are ``seen'' by the bath. In the dissipator given by Eq. \eqref{dissipator}, this renormalizes the coupling strengths $\Gamma_k^\pm$ to
\begin{equation}
    \Gamma_k^\pm = \sum_{m = -\infty}^{+\infty} P(m) \gamma_k^\pm (\omega_0 + m \omega_{\rm F}),
\end{equation}
with the coefficients $P(m)$ given by the Floquet theory representing the weight of the mth Floquet sideband (see Appendix~\ref{model: Hamronic modulation}). We assume that the hot and cold reservoirs have non-overlapping spectral densities, narrowly centered around the frequencies\[\omega_h=\omega_0+\omega_{\rm F},\qquad\omega_c=\omega_0-\omega_{\rm F},\]
respectively. Consequently, the bath spectra select only the Floquet sidebands $m=+1$ and $m=-1$, while the contributions from all other harmonics vanish. In particular, the central harmonic at $\omega_0$ does not contribute, such that $\gamma_{h,c}^{\pm}(\omega_0)=0$. Then, using $P(\pm 1) = J^2_1\left(\Omega/\omega_{\rm F}\right)$, where $J_1$ is a Bessel function of the first kind, we obtain
\begin{align}
\Gamma_k^\pm &= J^2_1\left(\frac{\Omega}{\omega_{\rm F}}\right) \gamma_k^\pm (\omega_k),
\end{align} 
The power and currents of the ``Floquet-inspired engine'' are given by
\begin{align}
P_{\rm Floquet}(t) &= \Tr[\dot{H}_{\rm long}(t) \rho(t)], \\ %= \dot \omega(t) p_1(t), \\
J_{\rm Floquet}^{h,c} (t) &= \Tr[\pazocal{L}^\dag_{h,c} {H}_{\rm long}] \rho(t)].
\end{align}
Assuming the diagonal form of the density matrix, the system evolves according to the master equation $\dot \rho = (\pazocal{L}^{\rm F}_h + \pazocal{L}^{\rm F}_c) [\rho]$. Then, in the long-time steady state, we obtain (see Appendix \ref{model: Hamronic modulation}):
\begin{multline}
P_{\rm Floquet} = \lim_{t \to \infty} P_{\rm Floquet}(t) \\ 
= J^2_1\left(\frac{\Omega}{\omega_{\rm F}}\right) 
\frac{\left( \omega_h-\omega_c\right)  \gamma_c^- \gamma_h^- 
\left(e^{-\beta_h \omega_h}-e^{-\beta_c \omega_c}\right)}
{(1+e^{-\beta_c \omega_c}) \gamma_c^- +(1+e^{-\beta_h \omega_h}) \gamma_h^-},
\end{multline}
and 
\begin{equation}
\eta = \lim_{t \to \infty} \frac{P_{\rm Floquet}(t)}{J_{\rm Floquet}^h(t)} = 1 - \frac{\omega_c}{\omega_h},
\end{equation}
\begin{comment}
where we put
\begin{align}
 \omega_h = \omega_0 + \omega_{\rm F},\quad \omega_c = \omega_0 - \omega_{\rm F}.
\end{align} 
\end{comment}

The corresponding characteristic time of the Floquet engine is
\begin{align}
T^*_{\rm Floquet}=\frac{W_{\rm discrete}}{P_{\rm Floquet}}
=\frac{\tau_h +\tau_c}{J^2_1\left(\frac{\Omega}{\omega_{\rm F}}\right)},
\end{align}
where we recall that $\tau_h$ and $\tau_c$ are the thermalization time constants given by Eq. \eqref{eq: thermalization timescales}.

In the same spirit as in the previous model, we characterize the operational performance of the engine through the dimensionless parameter $s_F=\omega_{\rm F}/\Omega$. The resulting dimensionless operational timescale takes the form
\begin{align}
\frac{T^*_{\rm Floquet}}{\tau_h + \tau_c}= J^{-2}_1\left(s_F^{-1}\right),
\end{align}
The parameter $s_F=\omega_F/\Omega$ allows us to distinguish two qualitatively
different regimes. For $s_F>1$, the argument of the Bessel function is small,
$s_F^{-1}<1$, and
\begin{equation}
J_1^2(s_F^{-1}) \simeq \frac{1}{4s_F^2}.
\end{equation}
Consequently, the effective system--bath coupling through the selected Floquet sidebands is suppressed as $s_F^{-2}$, and the characteristic operational time grows quadratically with $s_F$.\newline \newline
For $s_F<1$, the argument of the Bessel function becomes of order unity or larger. In this regime, $J_1(s_F^{-1})$ becomes nonmonotonic and eventually oscillatory, passing through a sequence of zeros. Since the narrow spectral densities considered here select only the Floquet sidebands $m=\pm1$, the
effective bath coupling vanishes whenever
\begin{equation}
J_1(s_F^{-1})=0.
\end{equation}
Thus, within the present spectrally selective model, periodic driving can completely suppress the corresponding relaxation channels. Such a drive-induced control of system--bath coupling is closely related to Floquet engineering of dissipation and to the suppression of relaxation in periodically
driven open quantum systems \cite{PhysRevE.87.012140,HausingerGrifoni2010}. As discussed in Sec.~\ref{section 3}, these zeros appear as divergences of the characteristic operational time.

\begin{table*}[!t]
\centering
\renewcommand{\arraystretch}{2.2}
\small

\resizebox{\textwidth}{!}{%
\begin{tabular}{|c|c|c|c|c|}
\hline

&
\textbf{Four-stroke}
&
\textbf{Floquet}
&
\textbf{Always-on}
&
\textbf{On-off}
\\
\hline

\multirow{2}{*}{$\displaystyle
\frac{T_\alpha^*}{\tau_h+\tau_c}
$}

&

$\displaystyle
\frac{\kappa}{2}
%\frac{\sigma}{1+\sigma^2}
\frac{(s_{\rm LT}+s_h+s_c)
\left(1-e^{-(\sigma s_h+s_c/\sigma)}\right)}
{
\left(1-e^{-\sigma s_h}\right)
\left(1-e^{-s_c/\sigma}\right)}
$

&

$\displaystyle
J_1^{-2}\left(s_F^{-1}\right)
$

&

$\displaystyle
1+\frac{s_\pi^2}{\pi^2}
$

&

$\displaystyle
\frac{\kappa}{2}
%\frac{\sigma}{1+\sigma^2}
\frac{(s_\pi+s_{hc})
\left(1-e^{-2s_{hc}/\kappa}\right)}
{\left(1-e^{-\sigma s_{hc}}\right)
\left(1-e^{-s_{hc}/\sigma}\right)}
$

\\
\cline{2-5}

&

$\displaystyle
s_{\rm LT}=\frac{t_{\rm LT}}{\sqrt{\tau_h\tau_c}},
\quad
s_h=\frac{t_h}{\sqrt{\tau_h\tau_c}},
\quad
s_c=\frac{t_c}{\sqrt{\tau_h\tau_c}}
$

&

$\displaystyle
s_F=\frac{\omega_{\rm F}}{\Omega}
$

&

$\displaystyle
s_\pi=\frac{t_\pi}{\sqrt{\tau_h\tau_c}}
$

&

$\displaystyle
s_\pi=\frac{t_\pi}{\sqrt{\tau_h\tau_c}},
\quad
s_{hc}=\frac{t_{hc}}{\sqrt{\tau_h\tau_c}}
$

\\
\hline
\end{tabular}%
}

\caption{
Comparison of the four quantum Otto engine models. The first row gives the dimensionless characteristic operational time \(T_\alpha^*/(\tau_h+\tau_c)\), while the second row defines the dimensionless parameters used in each implementation. Here, \(t_\pi=\pi/\Omega\),
\(\kappa=2\sqrt{\tau_h\tau_c}/(\tau_h+\tau_c)\), and \(\sigma=\sqrt{\tau_c/\tau_h}\).}
\label{tab:comparison}
\end{table*}

\subsection{Transverse coupling}\label{Transverse coupling}
Now, we turn to transverse coupling models. These are specifically characterized by the two qubits with constant frequencies $\omega_h$ and $\omega_c$, and exposed to external transverse driving. In particular, we propose the time-dependent Hamiltonian:  
\begin{align}\label{eq:driven Hamiltonian1}
{H}_{\rm tran}(t)={H}_{0}+ {V}(t)\,,
\end{align}
with
\begin{align}
{H}_{0} &= {H}_h + {H}_c = \omega_h \sigma^+_h \sigma^-_h +  \omega_c \sigma^+_c \sigma^-_c \\
{V}(t) &= \frac{\Omega}{2} \left(\sigma^-_h \sigma^+_c e^{ i \omega_d t} + h.c. \right)\, \label{eq:drive1}.
\end{align}
 $\sigma^{\pm}_{h,c}$ are ladder operators associated with the qubits, $\Omega$ the strength of the drive, and $\omega_d=\omega_h-\omega_c$ its frequency (at resonance).

 Furthermore, the qubits are locally coupled to the corresponding heat baths, with dissipators given by (see Eqs. \eqref{dissipator}  and \eqref{coupling_rates}):
 \begin{equation}
\pazocal{L}_{k} [\rho] = \gamma^+_{k} \pazocal{D}[\sigma_{k}^+]\rho + \gamma^-_{k} \pazocal{D}[\sigma_{k}^-]\rho.
 \end{equation}
 The local master equation is justified here when $\Omega \ll \omega_h,\,\omega_c$.

\subsubsection{Always-on model}\label{section: Always-on model}
In contrast to the stroke-based engines discussed previously, the present model does not possess distinct thermalization and work strokes. Instead, it operates autonomously in a nonequilibrium stationary regime maintained by the two baths and the external drive. The state of the system then evolves according to the master equation:
\begin{align} \label{eq:always_on_rotating_interaction}
\dot{\rho}(t) =-i\left[{H}_{\rm tran}(t),\rho(t) \right]+ \pazocal{L}_h\left[ \rho(t)\right]+\pazocal{L}_c\left[ \rho(t)\right]\,.
\end{align}
The time-dependence in Eq.~\eqref{eq:driven Hamiltonian1} can be removed at resonance by transforming to the interaction picture generated by $ H_0$. the interaction Hamiltonian becomes
\begin{equation}
 V_I = {V}(0) = \frac{\Omega}{2}\left( \sigma_h^-\sigma_c^+ + \sigma_h^+\sigma_c^-\right).
\label{eq:always_on_rotating_interaction}
\end{equation}
The long-time state is therefore stationary in the rotating frame, whereas in the laboratory frame it corresponds, in general, to a periodic steady state.

The interaction in Eq.~\eqref{eq:always_on_rotating_interaction} coherently couples the single-excitation states
\begin{equation}
 \ket{10}\longleftrightarrow\ket{01}.
\end{equation}
A net engine event consists of the hot bath exciting the hot qubit, followed by the drive-assisted transfer of this excitation to the cold qubit and its subsequent relaxation into the cold bath. During such an event, the system absorbs the energy $\omega_h$ from the hot bath and releases the energy $\omega_c$ into the cold bath.

Similarly to the previous model, the power and heat currents are equal to:
\begin{align}
P_{\rm always-on}(t) &=\Tr\left[\dot{{H}}_{\rm tran}(t) \rho(t) \right], \\
J_{\rm always-on}^{h,c} (t) &= \Tr[\pazocal{L}^\dag_{h,c} [{H}_{\rm tran}] \rho(t)].
\end{align}
In the long-time regime, the amount of power extracted reads (see Appendix~\ref{model: always-on} for calculations): 
\begin{align}
P_{\rm always-on} &= \lim_{t \to \infty} P_{\rm always-on }(t) \\
&= \frac{W_{\rm discrete}}{\left(\tau_h+\tau_c \right)\left( 1+\frac{1}{\tau_h \tau_c \Omega^2} \right)}
\end{align}
which depends only on the characteristic thermalization times and the coherent coupling strength. The Otto efficiency is then obtained:
\begin{equation}
\eta = \lim_{t \to \infty} \frac{P_{\rm always-on}(t)}{J_{\rm always-on}^h(t)} = 1 - \frac{\omega_c}{\omega_h}.
\end{equation}
The corresponding characteristic time is given by
\begin{align}
T^*_{\rm always-on}&= \frac{W_{\rm discrete}}{P_{\rm always-on}}=\big(\tau_h+\tau_c \big)\left(1+\frac{t_{\pi}^2}{\pi^2 \tau_h \tau_c} \right)
\end{align} 
where we introduced the duration of the $\pi$-pulse: $t_\pi = \pi/\Omega$.

The dimensionless characteristic time is obtained by introducing $s_\pi=t_\pi/\sqrt{\tau_h \tau_c}$, yielding
\begin{align}
 \frac{T^*_{\rm always-on}}{\tau_h + \tau_c} =1 + \frac{s_\pi^2}{\pi^2}.
\end{align}

\subsubsection{On-off model}\label{On-and-off model}
We finally consider the case in which the driving protocol in Eq.~\eqref{eq:drive1} alternates between two switching configurations, namely the ``on and off'' regimes. When the drive is turned on, the qubits are decoupled from the baths, and the driving field realizes the $\pi$-pulse, inducing a transition between two intermediate energy levels of the system, $\ket{01} \leftrightarrow \ket{10}$. This stage lasts $t_\pi = \pi/\Omega$. 
Subsequently, the drive is turned off, and each qubit re-thermalizes with its corresponding thermal bath, thereby approaching the original thermal population of qubits. By $t_{hc}$ we denote the duration of this thermalization process, as the ``hot'' and ``cold'' thermalize simultaneously here. 

The single $m$-th cycle of the process 
\begin{equation}\label{on-off stages}
    \rho_0^{(m)}
\xrightarrow{t_\pi}
\rho_1^{(m)} = {U}_{t_\pi} \rho_0^{(m)} {U}_{t_\pi}^\dag 
\xrightarrow{t_{hc}}
\rho_2^{(m)} = \Lambda_{t_{hc}}[\rho_1^{(m)}]
\end{equation}
where 
\begin{align}
{U}_{t_\pi} &= \pazocal{T} e^{-i \int_0^{t_\pi} {H}_{\rm tran} (s) ds}, \quad
 \Lambda_{t_{hc}} = e^{(\pazocal{L}_h + \pazocal{L}_c)t_{hc}}.
\end{align}
The work and heat exchanged in a single cycle are defined as:
\begin{align}
W_{\rm on-off}^{(m)} &= -\Tr[{H}_0 (\rho_1^{(m)} - \rho_0^{(m)})] \\
{Q^{h,c \ (m)}_{\rm on-off}} &= \Tr[{H}_{h,c} (\rho_1^{(m)} - \rho_0^{(m)})]
\end{align}
As before, we consider a stationary limit such that the power is given by (see Appendix~\ref{model: on-and-off} for more details): 
\begin{align}
P_{\rm on-off} &= \frac{1}{\tau_{\rm on-off}}\lim_{m \to \infty} W_{\rm on-off}^{(m)} \nonumber \\
&= \frac{\left(1 - e^{ -t_{hc} /\tau_h} \right) \left(1 - e^{ - t_{hc}/\tau_c} \right)}{(t_\pi + t_{hc}) (1 - e^{-t_{hc}\left(1/\tau_h +1/\tau_c \right)} )} W_{\rm discrete}
\end{align}
and the efficiency 
\begin{equation}
\eta = \lim_{m \to \infty} \frac{W_{\rm on-off}^{(m)}}{Q_{\rm on-off}^{h \ (m)}} = 1 - \frac{\omega_c}{\omega_h}.
\end{equation}
The corresponding characteristic time of the ``on and off engine'' is therefore
\begin{align}
T^*_{\rm on-off}&= \frac{W_{\rm discrete}}{P_{\rm on-off}} \\
&= \frac{(t_\pi+t_{hc})(1 - e^{-t_{hc}\left(1/\tau_h +1/\tau_c \right)})}{\left(1 - e^{ -t_{hc} /\tau_h} \right) \left(1 - e^{ - t_{hc}/\tau_c} \right)} 
\end{align}
Introducing the dimensionless thermalization duration $s_{hc} =t_{hc}/\sqrt{\tau_h\tau_c}$ and the dimensionless modulation timescale $s_\pi$ as defined in the previous model, the operational timescale can be recast in dimensionless form as
\begin{align}\label{eq:recast-on-off time}
 \frac{T^*_{\rm on-off}}{\tau_h + \tau_c} =\frac{\kappa}{2} \frac{( s_\pi+s_{hc}) \left(1 - e^{ -2 s_{hc}/\kappa} \right)}{\left(1 - e^{ -\sigma s_{hc}} \right)\left(1 - e^{ -s_{hc} / \sigma} \right)},
\end{align}
with $\kappa$ and $\sigma$ given by Eq. \eqref{kappa_and_sigma}

\begin{figure}[t] 
\centering
\includegraphics[width=0.9\linewidth]{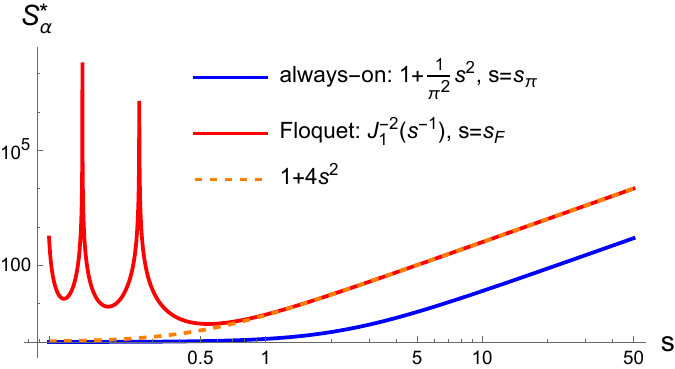}
\caption{Normalized characteristic operational times for the always-on and Floquet implementations as functions of their respective dimensionless control parameters. The blue curve shows the always-on result, $S_{\mathrm{always-on}}^*=1+s_\pi^2/\pi^2$, while the red curve shows $S_{\mathrm{Floquet}}^*=J_1^{-2}(s_F^{-1})$. The orange dashed line denotes the large-$s_F$ Floquet asymptote $1+4s_F^2$. The nonmonotonic behavior and divergences of the Floquet curve at small $s_F$ are discussed in the main text. Since $s_\pi$ and $s_F$ have different physical definitions, the figure compares their scaling behavior rather than their performance at fixed experimental resources.}
\label{fig:alwaysonfloquet}
\end{figure}

\begin{figure}[t] 
\centering
\includegraphics[width=0.9\linewidth]{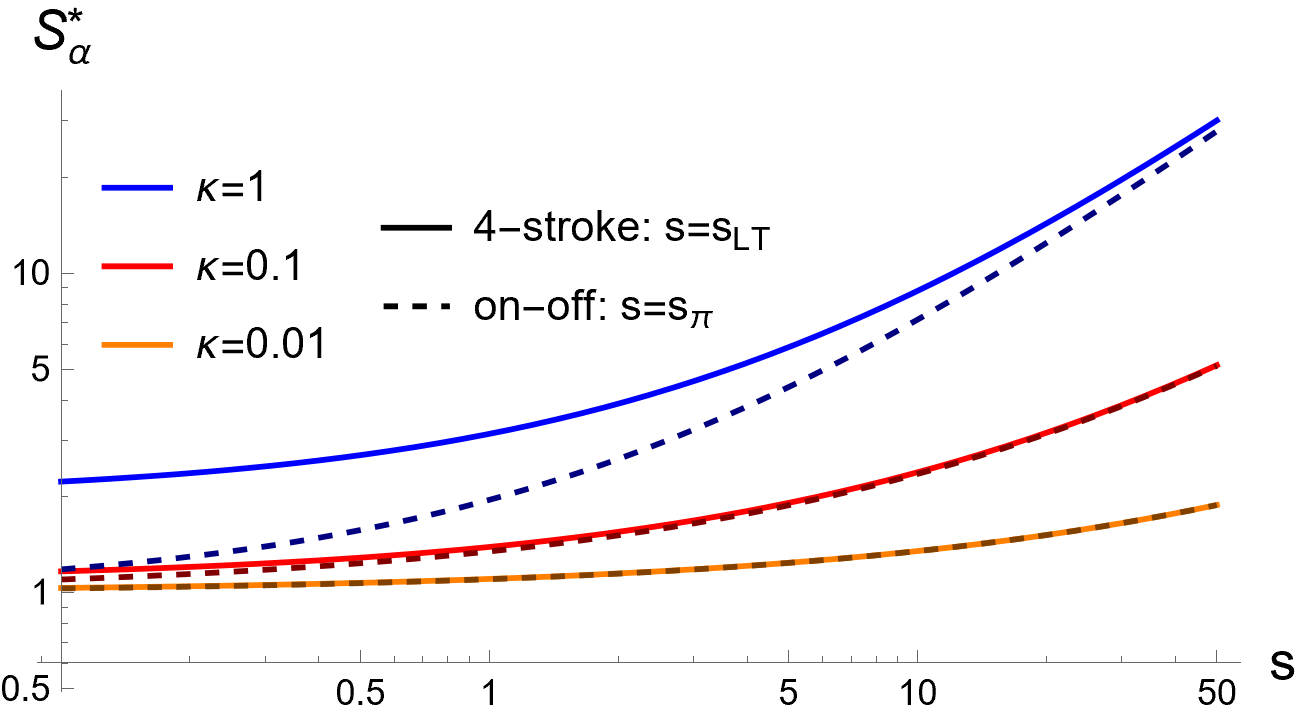}
\includegraphics[width=0.9\linewidth]{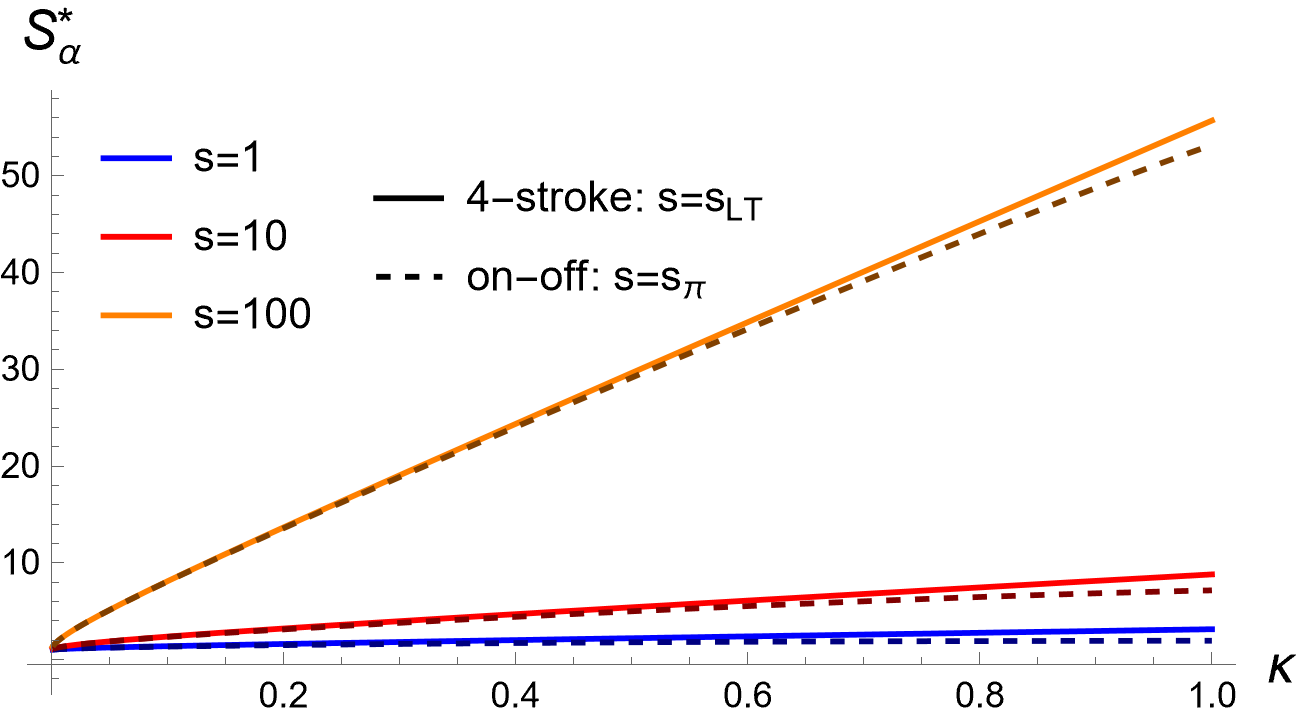}
\caption{Optimized normalized characteristic operational times of the four-stroke (solid curves) and on-off (dashed curves) implementations. The upper panel shows their dependence on the model-specific operation parameter for $\kappa=1$, $0.1$, and $0.01$. The lower panel shows their dependence on the relaxation-time asymmetry $\kappa$ for fixed operation parameters $s=1$, $10$, and $100$. Here $\kappa=1$ corresponds to equal relaxation times, whereas $\kappa\to0$ denotes the strongly asymmetric limit. Both implementations exhibit asymptotically linear growth with leading slope $\kappa/2$.}
\label{fig:fourstrokeonoff}
\end{figure}

\begin{figure}[t]
\centering
\includegraphics[width=0.9\linewidth]{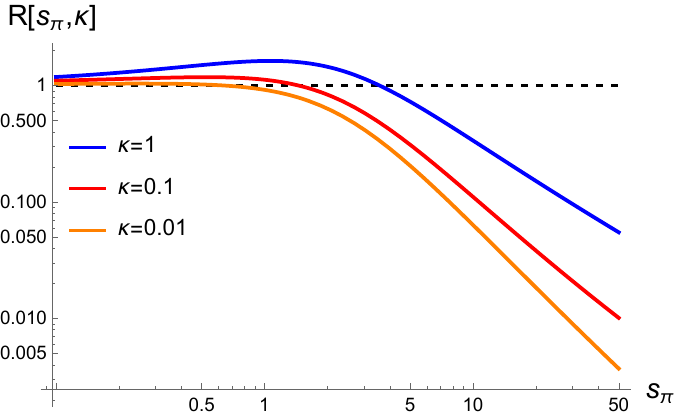}
\caption{Ratio \(R=S^*_{\rm on-off}(s_\pi,\kappa)/S^*_{\rm always-on}(s_\pi)\) as a function of the common operation parameter $s_\pi$ for different values of the relaxation-time asymmetry $\kappa$. The line $R=1$ (dashed) separates the regimes in which the always-on $(R>1)$ and on-off $(R<1)$ implementations have the shorter characteristic operational time.}
\label{fig:ratio-always-on-off}
\end{figure}

\section{Results and Discussion} \label{section 3}
We now compare the four dynamical implementations introduced in Secs.~\ref{Longitudinal coupling} and~\ref{Transverse coupling} using the characteristic operational time
\begin{equation}
T_\alpha^* = \frac{W_{\mathrm{discrete}}}{P_\alpha},
\end{equation}
where $\alpha$ labels the implementation. This quantity is the time required for an engine operating at power $P_\alpha$ to produce an amount of work equal to the work $W_{\mathrm{discrete}}$ of the fully thermalized discrete Otto cycle. Accordingly, a smaller $T_\alpha^*$ corresponds to a larger output power.

\subsection{Always-on and Floquet models}
The results obtained using the ``always-on'' and ``Floquet'' models are similar. For both models, the rescaled characteristic time depends solely on a single dimensionless parameter. Accordingly, we consider the following single-parameter functions:
\begin{equation}
S^*_{\rm always-on}(s_\pi) \equiv \frac{T^*_{\rm always-on}(s_\pi)}{\tau_h+\tau_c} = 1+\frac{s_\pi^2}{\pi^2},
\end{equation}
and
\begin{equation}
S^*_{\rm Floquet}(s_{\rm F}) \equiv \frac{T^*_{\rm Floquet}(s_{\rm F}^{-1})}{\tau_h+\tau_c} = J_1^{-2}(s_{\rm F}^{-1}).
\end{equation}

For large values of $s_{\rm F}$, we obtain
\begin{equation}
S^*_{\rm Floquet}(s_{\rm F}) = 1+4s_{\rm F}^2+O(s_{\rm F}^{-2}),
\end{equation}
which has the same quadratic scaling as the ``always-on'' model. For $s_F<1$, the Bessel-function dependence leads to a nonmonotonic characteristic time with a sequence of divergences. These divergences occur at the zeros of $J_1(s_F^{-1})$, where the weights of the $m=\pm1$ Floquet sidebands vanish. Because these are the only sidebands coupled to the spectrally narrow reservoirs in our model, the effective thermalization rates vanish at these points and the characteristic operational time diverges. Figure~\ref{fig:alwaysonfloquet} compares the always-on and Floquet implementations as functions of their respective dimensionless control parameters. However, this comparison establishes just a common quadratic asymptotic scaling for the always-on and Floquet implementations rather than providing a point-by-point performance comparison. In particular, $s_\pi$ and $s_{\rm F}$ describe different physical control parameters, so equal numerical values do not represent identical experimental resources.

\subsection{Four-stroke and on-off models}
The four-stroke and on-off implementations form a second natural class: in both protocols, work extraction alternates with finite thermalization stages. Their characteristic times, however, initially depend on several dimensionless durations, and an additional optimization is therefore required before their scaling behavior can be compared.

Let us first consider the fully asymmetric limit, in which the parameter
\begin{equation}
\sigma=\sqrt{\frac{\tau_c}{\tau_h}}
\end{equation}
either approaches zero, $\sigma\to0$, meaning that equilibration with the cold bath is much faster than equilibration with the hot bath, or approaches infinity, $\sigma\to\infty$, corresponding to the opposite limit. In both cases, the asymmetry parameter
\begin{equation}
\kappa = \frac{2\sqrt{\tau_h\tau_c}}{\tau_h+\tau_c} = 2\left(\sigma+\frac{1}{\sigma}\right)^{-1}
\end{equation}
approaches zero, $\kappa\to0$. For the ``four-stroke'' model, we then obtain two solutions:
\begin{equation}
\lim_{\sigma\to0} \frac{T_{\rm 4-stroke}^*}{\tau_h+\tau_c} = 1+\frac{t_{\rm LT}}{t_h}+\frac{t_c}{t_h},
\end{equation}
and
\begin{equation}
\lim_{\sigma\to\infty} \frac{T_{\rm 4-stroke}^*}{\tau_h+\tau_c} = 1+\frac{t_{\rm LT}}{t_c}+\frac{t_h}{t_c},
\end{equation}
whereas the ``on-off'' model gives, in both cases,
\begin{equation}
\lim_{\kappa\to0} \frac{T_{\rm on-off}^*}{\tau_h+\tau_c} = 1+\frac{t_\pi}{t_{hc}}.
\end{equation}
In this limit, we obtain very simple expressions with linear scaling governed by the newly introduced parameters. Interestingly, the common lower limit of the characteristic times $T_\alpha^*$ for these models is again given by $\tau_h+\tau_c$, which is reached when the thermalization time ($t_h$, $t_c$, or $t_{hc}$) tends to infinity. This counterintuitive result—namely, that maximal power is achieved in the limit of infinite stroke duration—can be understood by noting that one of the corresponding characteristic thermalization times, either $\tau_h$ or $\tau_c$, is not fixed but also tends to infinity in this case.

To compare the models for $\kappa>0$, we introduce an additional optimization by defining the following functions:
\begin{align}
 S^*_{\rm 4-stroke}(s_{\rm LT},\kappa) &\equiv \min_{s_h,s_c} \frac{ T^*_{\rm 4-stroke}(s_{\rm LT},s_h,s_c,\sigma)}{\tau_h+\tau_c}, \\
S^*_{\rm on-off}(s_\pi,\kappa) &\equiv \min_{s_{hc}} \frac{T^*_{\rm on-off}(s_\pi,s_{hc},\sigma)}{\tau_h+\tau_c}.
\end{align}

Notice that, whereas $T^*_{\rm always-on}$ depends directly on $\sigma=\sqrt{\tau_c/\tau_h}\in(0,\infty)$, the optimized functions $S^*_{\rm 4-stroke}$ and $S^*_{\rm on-off}$ are invariant under the transformation $\sigma\to1/\sigma$. Therefore, it is sufficient to consider only one branch, i.e., $\sigma\in(0,1]$ or $\sigma\in[1,\infty)$. For this reason, we propose using $\kappa$ rather than $\sigma$ as the asymmetry parameter, since it is symmetric with respect to $\tau_h$ and $\tau_c$.

The optimization reduces the dependence of both protocols to the operation parameter and the relaxation-time asymmetry $\kappa$. Their behavior can then be characterized analytically in the limits of fast and slow work strokes. For $s_{\rm LT}\ll1$, the optimized four-stroke characteristic time is
\begin{multline}
S^*_{\rm 4-stroke}(s_{\rm LT},\kappa) = 1+\kappa + \left(\frac{9}{32}\kappa(1+\kappa)^2 \right)^{1/3}
s_{\rm LT}^{2/3}\\
+O(s_{\rm LT}^{4/3}),
\end{multline}
whereas the corresponding expansion for the on-off protocol is
\begin{multline}
S^*_{\rm on-off}(s_\pi,\kappa) = 1 + \left(\frac{9}{16}\right)^{1/3} s_\pi^{2/3} + O(s_\pi^{4/3}).
\end{multline}
These expressions show that the two protocols have different limiting values in the fast-operation regime. The four-stroke implementation approaches $1+\kappa$, whereas the on-off implementation approaches unity. In particular, for equal relaxation times, $\kappa=1$, their respective limiting values are $2$ and $1$. This difference reflects the fact that the four-stroke engine requires two separately optimized thermalization strokes, whereas both qubits thermalize simultaneously during the thermalization stage of the on-off protocol. The $s_\alpha^{2/3}$ scaling arises from the optimization of the competition between the duration of thermal contact and the reduction in extracted work caused by incomplete thermalization.

In the opposite limit of slow operation, the leading behavior becomes
\begin{equation}
S^*_{\rm 4-stroke}(s_{\rm LT},\kappa) \sim \frac{\kappa}{2}s_{\rm LT}, \quad
S^*_{\rm on-off}(s_\pi,\kappa) \sim \frac{\kappa}{2}s_\pi.
\end{equation}
Thus, despite their different structures and fast-operation limits, both optimized protocols exhibit the same asymptotically linear dependence, with a slope determined by the relaxation-time asymmetry. In this regime, the duration of the work stroke dominates the characteristic time, whereas the optimized thermalization contribution becomes subleading.

%Moreover, in the strongly asymmetric limit, $\kappa\ll1$, the optimized characteristic times take the form
Moreover, in the strongly asymmetric regime, $\kappa\ll1$, at fixed operation parameter—or, more generally, for $\kappa s_\alpha \ll 1$—the optimized characteristic times take the form
\begin{align}
S^*_{\rm 4-stroke}(s_{\rm LT},\kappa) &= 1+\sqrt{\kappa s_{\rm LT}} +\frac{\kappa}{3}s_{\rm LT}
+\dots, \\
S^*_{\rm on-off}(s_\pi,\kappa) &= 1+\sqrt{\kappa s_\pi} +\frac{\kappa}{3}s_\pi +\dots.
\end{align}
%Thus, in the limit $\kappa\to0$, both optimized characteristic times approach the same limiting value,
Thus, at fixed $s_{\mathrm{LT}}$ and $s_\pi$, both optimized characteristic times approach the same limiting value, as $\kappa \to 0$,
\begin{equation}
S^*_{\rm 4-stroke},S^*_{\rm on-off}\longrightarrow 1.
\end{equation}
Moreover, when the two protocols are compared at the same value of their respective dimensionless operation parameters $s_{\rm LT}=s_\pi$, their asymptotic expansions coincide with the orders shown above. This demonstrates that the distinction between sequential thermalization in the four-stroke protocol and simultaneous thermalization in the on-off protocol becomes progressively less important as the relaxation times become strongly asymmetric. In this regime, the slower bath sets the dominant thermalization timescale and therefore acts as the main bottleneck for both implementations, while protocol-specific differences contribute only through subleading corrections. This convergence is also visible in Fig.~\ref{fig:fourstrokeonoff}: as $\kappa$ decreases, the optimized characteristic times of the four-stroke and on-off engines approach one another, consistently with the asymptotic expressions above.

%\newpage
\subsection{Always-on and on-off}
Although the characteristic operational time provides a common measure of engine speed, the four implementations cannot in general be compared point by point, since their dimensionless control parameters have different physical meanings. An exception is given by the always-on and on-off models, which share the same transverse coupling and therefore the same parameter $s_\pi$, allowing a direct comparison of their characteristic times.

Fig.~\ref{fig:ratio-always-on-off} shows  on the log-log scale the ratio \(R(s_\pi,\kappa)=S^*_{\rm on-off}(s_\pi,\kappa)/S^*_{\rm always-on}(s_\pi)\), where $R>1$ indicates that the always-on engine is faster, while $R<1$ indicates that the on-off engine has the shorter characteristic time. The comparison shows that neither implementation is universally faster. At small and intermediate values of $s_\pi$, the ordering depends on both $s_\pi$ and $\kappa$, whereas at sufficiently large $s_\pi$ the on-off engine becomes faster for all displayed values of $\kappa$. This crossover reflects the linear asymptotic growth of $S_{\mathrm{on-off}}^*$, in contrast to the quadratic growth of $S_{\mathrm{always-on}}^*$.

%\newpage
\section{SUMMARY and outlook}\label{sec:summary}
In this work, we compared four dynamical implementations of the quantum Otto cycle: the four-stroke and Floquet engines, based on a single-qubit working medium, and the always-on and on-off engines, based on two coupled qubits. To place these implementations on a common thermodynamic footing, we introduced the characteristic operational time through the factorization in Eq.~(\ref{power_factorization}). Equivalently, $T_\alpha^*=W_{\mathrm{discrete}}/P_\alpha$ represents the effective time required for an engine operating at power $P_\alpha$ to produce an amount of work equal to that of the fully thermalized discrete Otto cycle. This quantity accounts for both the physical duration of the protocol and the reduction in work caused by incomplete thermalization. It therefore provides a more informative measure of operational speed than the cycle duration alone and remains applicable to implementations without a conventional stroke-based cycle.

Our results reveal two distinct asymptotic classes. After optimizing the thermalization durations, the characteristic times of the four-stroke and on-off protocols grow linearly with their respective dimensionless operation parameters. By contrast, the always-on and Floquet implementations exhibit quadratic growth in the corresponding slow-transfer or high-frequency regimes, respectively. We also showed that the asymmetry between the hot- and cold-qubit relaxation times affects the optimized four-stroke and on-off protocols. In the strongly asymmetric limit, their normalized characteristic times approach one another because the slower relaxation process becomes the dominant operational bottleneck. 

These results characterize the scaling of each implementation with its natural control parameter, but they do not generally constitute a point-by-point performance ranking under identical experimental resources, since the dimensionless parameters of the different models have different physical meanings. An important exception is provided by the always-on and on-off implementations, which share the same transverse coupling and therefore the same control parameter $s_\pi$. Their direct comparison shows that neither implementation is universally faster: the ordering depends on both $s_\pi$ and the relaxation-time asymmetry $\kappa$. Nevertheless, the on-off protocol becomes faster at sufficiently large $s_\pi$, owing to its linear asymptotic growth, in contrast to the quadratic growth of the always-on characteristic time. 

A broader experimental comparison would require additional common constraints, including fixed coupling strengths, admissible control bandwidths, dissipation rates, and the energetic costs of driving and switching the interactions. Extending the present analysis to include such control costs, nonideal work strokes, decoherence, and non-Markovian reservoirs would provide a more complete assessment of the practical advantages of the different implementations. The characteristic operational time introduced here offers a unified starting point for such comparisons and may help guide the selection of protocols for future experimental quantum heat engines.

\begin{acknowledgments}
The authors thank Micha{\l} Horodecki for helpful discussions. I.H.N.N. acknowledges financial support from the project ``International Centre for Theory of Quantum Technologies 2.0: R\&D Industrial and Experimental Phase'' (No.~FENG.02.01-IP.05-0006/23), funded under the European Funds for a Smart Economy 2021--2027 (FENG) programme. M. Ł. acknowledges the support by the Polish National Science Centre Grant OPUS-21 (No. 2021/41/B/ST2/03207).
\end{acknowledgments}

\bibliographystyle{apsrev4-2}
\bibliography{references_PRB_cleaned}

\newpage 
\clearpage
\onecolumngrid

\appendix
\section{Models}

\subsection{Dynamical four-stroke engine (longitudinal coupling)} \label{model: 4-stroke engine}
Our first dynamical model of the Otto engine is a finite-time implementation of the conventional four-stroke cycle. In this model, a qubit acts as a working medium and is successively coupled to hot and cold thermal reservoirs with inverse temperatures $\beta_h$ and $\beta_c$, respectively. The cycle consists of two isochoric strokes, during which heat is exchanged with one of the reservoirs, and two isolated level-transformation strokes, during which the qubit transition frequency is varied.

We consider the (periodic) longitudinal modulation $\omega(t)$ introduced in Eq.~\eqref{logitudinal_modulation}, satisfying 
\begin{align}
\omega(t)=\omega(t+\tau_{\rm 4-stroke}),
\end{align}
where 
\begin{align} 
\tau_{\rm 4-stroke}=t_h+t_{\rm LT,1}+t_c+t_{\rm LT,2}
\end{align}
is the period of one complete cycle. This modulation is slow compared to the thermalization timescales $\tau_h$ and $\tau_c$. This condition is necessary to write the full dynamics of the system as 
\begin{align}
\dot{\rho}(t) =-i \left[{H}_{\rm long}(t),\, \rho(t) \right] + \sum_{k=h,c} \chi_k(t)\pazocal{L}_{k}[\omega(t)]\rho(t).
\end{align}
where $\chi_k(t)$ is a switching function equal to unity when the system is coupled to reservoir $k$ and zero otherwise. In particular, both switching functions vanish during the isolated level-transformation strokes. The dissipator associated with reservoir $k$ is
\begin{equation}
 \pazocal{L}_{k}[\omega(t)]\rho =  \gamma_k^+(\omega(t)) \pazocal{D}[\sigma_+]\rho + \gamma_k^-(\omega(t)) \pazocal{D}[\sigma_-]\rho, 
\end{equation}
with 
\begin{align}
\pazocal{D}\left[ A \right]\rho=A \rho A^{\dagger}-\frac{1}{2} \left\{ A^{\dagger}A,\,\rho\right\}
\end{align}
During the hot and cold isochoric strokes, the transition frequency is held fixed at $\omega_h$ and $\omega_c$, respectively. 

The excited-state population
\begin{align}
p(t)=\bra{1}\rho(t)\ket{1}
\end{align}
then satisfies
\begin{align}
\dot p(t) = \pazocal{L}_{k}(\omega_k)p(t) = \Big(p^{\rm eq}_{k}-p(t)\Big)/\tau_{k}, \quad 1/\tau_k=\gamma^+_k(\omega_k)+\gamma^-_k(\omega_k),
\end{align}
where $\tau_k\,\,(k=h,c)$ is the characteristic thermalization timescale. $p^{\rm eq}_{k}$ is the excited-state population of the Gibbs state associated with reservoir $k$. Its solution is
\begin{align}
p(t)=e^{-\Gamma_k t}p(0)+ \big(1-e^{-\Gamma_k t} \big) p^{\rm eq}_{k},\quad \text{where} \quad \Gamma_k=1/\tau_{k}.
\end{align}
For states diagonal in the energy basis, this equation completely determines the state during each isochoric stroke.
\newline
\newline
In the first cycle, the state transformation can be represented as
\begin{align}\label{state evoluyion for ottoc cycle}
\rho_0^{(1)}
&\xrightarrow{t_h}
\rho_1^{(1)} =  \Lambda^h_{t_{h}}[\rho_0^{(1)}]
\xrightarrow{t_{\rm LT,1}} 
\rho_1^{(1)} \xrightarrow{t_{c}} \rho_2^{(1)} = \Lambda^c_{t_{c}} [\rho_1^{(1)}] \xrightarrow{t_{\rm LT,2}} \rho_2^{(1)}
\end{align}
where $\Lambda^{k}_{t} = e^{\pazocal{L}_{k}(\omega_k) t}$. Because the Hamiltonian remains diagonal in the fixed basis $\left\{\ket{0},\ket{1} \right\}$, the level-transformation strokes leave the populations unchanged. This does not require infinitely slow driving; it follows from
\begin{align}
 \left[\hat{H}_{\rm long}(t),\, \hat{H}_{\rm long}(t')\right]=0.
\end{align}

In the following subsections, we analyze each of the four strokes of the engine following the same framework introduced in Sec.~\ref{section: Discrete fours-stroke engine}. Each stroke corresponds to the standard expansion and compression protocols commonly considered in the literature. Together, these strokes form a complete thermodynamic cycle. We denote by $Q^{\alpha \ (m)}_{\rm 4-stroke}$ and $ W_{\rm 4-stroke}^{\alpha \ (m)}$ ($\alpha=h,c$) the heat and work exchanged during the $m^{\rm th}$ cycle, respectively. We first analyze the single-cycle regime and subsequently consider the $m^{\rm th}$ cycle, in the asymptotic limit of large $m$.

\subsubsection{Hot isochoric stroke (heating, duration $t_h$)}
The system is brought into contact with the hot bath, while the Hamiltonian is fixed at ${H}_h=\omega_h\ket{1}\bra{1}$. The heat exchanged $Q^{h \ (1)}_{\rm 4-stroke}$ between the system and the hot bath as the former's state evolves from $\rho(0)$ to $\rho(t_h)$ is given by
\begin{align}
Q^{h \ (1)}_{\rm 4-stroke}&=\Tr\Big[{H}_h \big(\rho_1^{(1)}-\rho_0^{(1)}\big)\Big] =\omega_h\delta p^{(1)}
\end{align}
where $\delta p^{(1)}= p_1^{(1)}- p_0^{(1)}$  with $p_j^{(1)}= \bra{1} \rho_j^{(1)} \ket{1}\,\,(i=1,2)$
\subsubsection{Adiabatic expansion (unitary stroke, duration $t_{\rm LT,1}$)}
We disconnect the hot bath and the Hamiltonian changes from ${H}_h=\omega_h\ket{1}\bra{1}$ to ${H}_c=\omega_c\ket{1}\bra{1}$. Since the system evolves unitarily, its state does not change, and the work done by the system is given by
\begin{align}
W_{\rm 4-stroke}^{h \ (1)}&=\Tr\Big[{H}_h\rho_1^{(1)}\Big]-\Tr\Big[{H}_c\rho_1^{(1)}\Big] \nonumber 
\\&=(\omega_h-\omega_c) p_1^{(1)} 
\end{align}
\subsubsection{Cold isochoric stroke (cooling, duration $t_c$)}
Now, the system interacts with the cold bath while the Hamiltonian is maintained at ${H}_c$. The population partially thermalizes toward the cold Gibbs state, exchanging heat 
$Q^{c \ (1)}_{\rm 4-stroke}$ given by
\begin{align}
Q^{c \ (1)}_{\rm 4-stroke}&=\Tr\Big[{H}_c \big(\rho_2^{(1)}-\rho_1^{(1)}\big)\Big] =-\omega_c\delta p^{(1)}
\end{align}

\subsubsection{Adiabatic compression (unitary stroke, duration $t_{\rm LT,2}$)}
The cold bath is removed and the Hamiltonian is changed from ${H}_c$ to ${H}_h$. The system evolves unitarily, and the energy change in this stroke corresponds to work
\begin{align}
W_{\rm 4-stroke}^{c \ (1)}&=\Tr\Big[{H}_c\rho_2^{(1)}\Big]-\Tr\Big[{H}_h\rho_2^{(1)}\Big] \nonumber \\
&=-(\omega_h-\omega_c) p_2^{(1)}
\end{align}
At times $t_h$ and $t_h+t_c$, the excited population has forms
\begin{align}\
p_1^{(1)}&=e^{-\Gamma_h t_h}p_0^{(1)} +\big(1-e^{-\Gamma_h t_h} \big) p^{\rm eq}_{h} \label{eq010} \\
p_2^{(1)}&=e^{-\Gamma_c t_c}p_1^{(1)} +\big(1-e^{-\Gamma_c t_c} \big) p^{\rm eq}_{c} \label{eq011}.
\end{align}
Plugging \eqref{eq010} into \eqref{eq011} yields the following:
\begin{align}\label{eq012}
p_2^{(1)}=A + Bp_0^{(1)},
\end{align}
with \begin{align}
A&= \big(1-e^{-\Gamma_c t_c} \big) p^{\rm eq}_{c} + e^{-\Gamma_c t_c}\big(1-e^{-\Gamma_h t_h} \big) p^{\rm eq}_{h}\,, \\
B&=e^{-\big(\Gamma_h t_h+\Gamma_c t_c\big)}\,.
\end{align}
Eq. \eqref{eq012} connects the population of the excited state at the beginning and the end of the same cycle.

\subsubsection{Asymptotic case}
Interestingly, in the m-th cycle, the state of the system transforms as:
\begin{align}
    \rho_0^{(m)}
&\xrightarrow{t_h}
\rho_1^{(m)} =  \Lambda^h_{t_{h}}[\rho_0^{(m)}]
\xrightarrow{t_{\rm LT,1}} 
\rho_1^{(m)} \xrightarrow{t_{c}} \rho_2^{(m)} = \Lambda^c_{t_{c}} [\rho_1^{(m)}] \xrightarrow{t_{\rm LT,2}} \rho_2^{(m)}
\end{align}

Following Eq. \eqref{eq012},we obtain a recurrence relation for the excited-state population given by
\begin{align}\label{eq015}
p_2^{(m)}=A + Bp_0^{(m)}
\end{align}
Using this generalization, one could evaluate the work and heat for this cycle. For the first stroke, the initial excited population is $p_n$ and the population at time $t_h$ corresponds to
\begin{align}\label{eq016}
p_1^{(m)}=e^{-\Gamma_h t_h}p_0^{(m)} +\big(1-e^{-\Gamma_h t_h} \big) p^{\rm eq}_{h}\,.
\end{align}
This leads to a heat of the form
\begin{align}
Q^{h \ (m)}_{\rm 4-stroke}&=\Tr\Big[{H}_h \big(\rho_1^{(m)}-\rho_0^{(m)}\big)\Big] =\omega_h\delta p^{(m)}
\end{align}
where $\delta p^{(m)}= p_1^{(m)}- p_0^{(m)}$  with $p_j^{(m)}= \bra{1} \rho_j^{(m)} \ket{1}\,\,(j=0,1,2)$.
For the other strokes, following the same principle, we obtain the following
\begin{align}
W_{\rm 4-stroke}^{h \ (m)}&=(\omega_h-\omega_c) p_1^{(m)}\\
Q^{c \ (m)}_{\rm 4-stroke}&=-\omega_c\delta p^{(m)}\\
W_{\rm 4-stroke}^{c \ (m)}&=-(\omega_h-\omega_c) p_2^{(m)}
\end{align}
The total work extracted from the system at the end of the $m^{th}$ cycle is given by
\begin{align}
W_{\rm 4-stroke}^{\ (m)}&=W_{\rm 4-stroke}^{h \ (m)} + W_{\rm 4-stroke}^{c \ (m)}= (\omega_h-\omega_c)\delta p^{(m)}  \label{eq023}
\end{align}
From Eq. \eqref{eq015}, we obtain a relation between $ p_2^{(m)}$ and $ p_0^{(1)}$ of the form
\begin{align}\label{eq024}
 p_2^{(m)}=A\Big(\frac{1-B^m}{1-B}\Big)+B^m  p_0^{(1)}
\end{align}
Plugging \eqref{eq024} and \eqref{eq015} into \eqref{eq023} and taking $m \to \infty$ ($B^m\to 0$), we obtain an asymptotic expression of \eqref{eq023} of the form
\begin{align}
W_{\rm 4-stroke}^{\rm asy} & = \lim_{m \to \infty}W_{\rm 4-stroke}^{\ (m)} = \Bigg[\frac{\left(1 - e^{-t_h\Gamma_h} \right) \left( 1 - e^{ -t_c\Gamma_c} \right) }{1 - e^{-({ t_h\Gamma_h} +  t_c\Gamma_c)}}\Bigg]  (\omega_h-\omega_c)\delta p,
\end{align}
with $\delta p$ given in Eq.~\eqref{eq:deltap}. The power of the engine in the asymptotic limit is given by:
\begin{align}\label{power_4stroke}
    &P_{\rm 4-stroke} = \frac{W_{\rm 4-stroke}^{\rm asy}}{\tau_{\rm 4-stroke}}  
    = \Bigg[\frac{\left(1 - e^{ -t_h\tau_h^{-1}} \right) \left( 1 - e^{ -t_c\tau_c^{-1}} \right) }{(t_h+t_c+ t_{\rm LT})(1 - e^{-({ t_h\tau_h^{-1}} +  t_c\tau_c^{-1})})}\Bigg] W_{\rm discrete},
\end{align}
where $t_{\rm LT}=t_{\rm LT,1}+t_{\rm LT,2}$

\subsubsection{Operational time}
From the asymptotic power in Eq. \eqref{power_4stroke},  we define the characteristic operational time 
\begin{align}
T^*_{\rm 4-stroke} =\frac{W_{\rm discrete}}{P_{\rm 4-stroke}}=\frac{(t_h+t_c+t_{\rm LT})(1 - e^{-({ t_h/\tau_h} +  t_c/\tau_c)})}{\left(1 - e^{ -t_h/\tau_h} \right) \left( 1 - e^{ -t_c/\tau_c} \right) }.
\end{align}
This quantity represents the effective time required, at the asymptotic cycle-averaged power, to produce an amount of work equal to $W_{\rm discrete}$, the work extracted during a fully thermalized discrete Otto cycle. It is therefore a comparative performance timescale rather than a transient first-passage time. A smaller $T^*_{\rm 4-stroke}$ corresponds to a larger asymptotic output power.

This time depends on several timescales associated with thermalization and level transformations. To facilitate the analysis, we recast this expression in terms of dimensionless variables:
\begin{align}
s_{\rm LT}=\frac{t_{\rm LT}}{\sqrt{\tau_h \tau_c}},\quad s_h=\frac{t_h}{\sqrt{\tau_h \tau_c}},\quad s_c=\frac{t_c}{\sqrt{\tau_h \tau_c}},
\end{align}
such that
\begin{align}\label{eq1:recast-four-stroke time}
\frac{T^*_{\rm 4-stroke}}{\tau_h + \tau_c}  =\frac{\kappa}{2}\frac{(s_{\rm LT}+s_h+s_c)
\left(1-e^{-(\sigma s_h+s_c/\sigma)}\right)}{\left(1-e^{-\sigma s_h}\right)\left(1-e^{-s_c/\sigma}\right)}
\end{align}
where $\sigma=\kappa^{-1}+\sqrt{\kappa^{-2}-1}$ and the asymmetry parameter $\kappa$ is given in Eq. \eqref{kappa}. We then minimize the above expression with respect to the dimensionless thermalization variables $s_h$ and $s_c$, thereby defining the optimal dimensionless operational timescale
\begin{align}
S^*_{\rm 4-stroke}(s_{\rm LT})&= \underset{s_h,\,s_c \geq 0}{\rm min}\left( \frac{T^*_{\rm 4-stroke}}{\tau_h + \tau_c}\right) 
\end{align}
Although a closed analytical expression for \(S^*_{\rm 4-stroke}(s_{\rm LT})\) is not available, several useful properties can be established. First, we can derive a bound to that minimum for very small values of $s_h$ and $s_c$ using the Taylor expansion:
\begin{align}
S^*_{\rm 4-stroke}(s_{\rm LT})&=\frac{\kappa}{2}\underset{s_h,\,s_c \geq 0}{\rm min}\frac{(s_{\rm LT}+s_h+s_c)
\left(1-e^{-(\sigma s_h+s_c/\sigma)}\right)}{\left(1-e^{-\sigma s_h}\right)\left(1-e^{-s_c/\sigma}\right)}\nonumber \\
& \approx \frac{\kappa}{2}\Big[s_{\rm LT} \left(\frac{1}{a}+\frac{1}{b}\right)+\frac{1}{\sigma}+\sigma +\frac{a}{\sigma b}+\frac{\sigma b}{a} \Big]
\end{align}
where $a=\sigma s_h$, $b=\frac{s_c}{\sigma}$. Using the inequality $\frac{a}{\sigma b}+\frac{\sigma b}{a} \geq 2$ and the fact that  $\frac{1}{\sigma}+\sigma=\frac{2}{\kappa}$, we obtain 
\begin{align}
S^*_{\rm 4-stroke}(s_{\rm LT}) \geq 1 +\kappa+\frac{\kappa}{2}s_{\rm LT}\left(\frac{1}{a} +\frac{1}{b} \right)
\end{align}
For $s_{\rm LT}=0$, we obtain the bound 
\begin{align}
S^{*}_{\rm 4-stroke}(0)&= 1+\kappa,
\end{align}

\begin{figure}[t!]
\centering
\includegraphics[width=.8\linewidth]{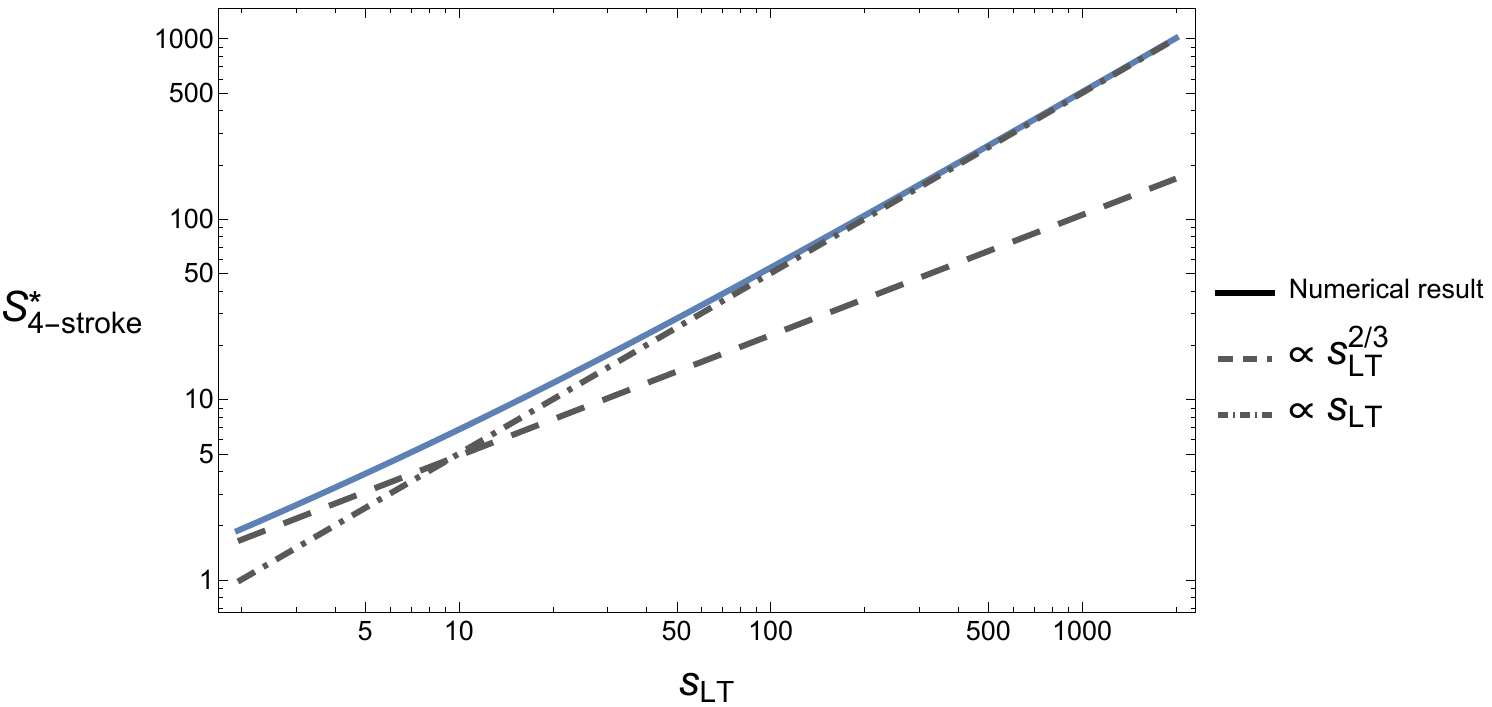}
\caption{
Log-log plot of the optimized dimensionless operational timescale
\(S^*_{\rm 4\text{-}stroke}\) as a function of the dimensionless parameter
\(s_{\rm LT}\) for $\kappa=1$. The solid curve corresponds to the numerical optimization, while the dashed and dot-dashed curves represent the asymptotic scalings \(S^*_{\rm 4\text{-}stroke}-(1+\kappa)\propto +s_{\rm LT}^{2/3}\) and \(S^*_{\rm 4\text{-}stroke}\propto s_{\rm LT}\), respectively. The logarithmic scales emphasize the crossover between the two asymptotic power-law regimes. }
\label{fig:asymptotic-fourstroke}
\end{figure}

For finite values of the level-transformation parameter, the optimized operational timescale exhibits two distinct asymptotic regimes,
\begin{align}
S^*_{\rm 4\text{-}stroke}(s_{\rm LT})-(1+\kappa)
\propto
s_{\rm LT}^{2/3},
\qquad
s_{\rm LT}\ll1,
\end{align}
and
\begin{align}
S^*_{\rm 4\text{-}stroke}(s_{\rm LT})
\propto
s_{\rm LT},
\qquad
s_{\rm LT}\gg1.
\end{align}
These asymptotic behaviors are illustrated in Fig.~\ref{fig:asymptotic-fourstroke}, where the dimensionless operational timescale is shown on logarithmic scales for both axes. The log-log representation highlights the two power-law regimes, with the numerical optimization approaching slopes of \(2/3\) and \(1\) in the small- and large-\(s_{\rm LT}\) limits, respectively.

\subsection{Harmonically modulated engine (longitudinal coupling)}\label{model: Hamronic modulation}
Here, we study an engine model similar to the one discussed above. The external drive is still periodic but has the following form:

%We now turn to a case similar to the one just above, consisting of a quantum Otto engine, where the working medium has the same Hamiltonian as \eqref{eq:Hamiltonian4strokes}. However, instead of switching the frequency from $\omega_h$ to $\omega_c$ and vise versa, we use a sinusoidal time-dependent field corresponding to an external modulation of the form 
\begin{equation}
\omega(t) = \omega_0 + \Omega \sin(\omega_{\rm F} t)\,,
\end{equation}
where $\omega_0$ is the mean transition frequency, $\Omega$ is the modulation amplitude, and $\omega_{\rm F}$ is the modulation frequency. The period of the drive is given by:
\begin{align}
\tau_{\rm F} = \frac{2\pi}{\omega_{\rm F}}
\end{align}
 In the regime of fast oscillations, $\omega_{\rm F} \gg 1/\tau_{h,c}$, the system can be described using Floquet theory \cite{ASENS_1883_2_12__47_0, PhysRev.138.B979, PhysRevA.7.2203}. In particular, this model was studied in \cite{PhysRevE.87.012140} to design a universal heat engine. In the same regime, we observe the creation of new harmonics of the form $\omega_m=\omega_0+m\omega_{\rm F}$ ($m=0,\pm1,\pm2,\dots$).

\subsubsection{Master equation}
The dynamics of the qubit are described by a Markovian master equation. We consider bosonic reservoirs weakly coupled to the system and, owing to the periodic modulation of the Hamiltonian, employ a Floquet expansion of the system operators. Furthermore, we assume a weak driving amplitude, $(\Omega\ll\omega_{\rm F})$, such that only the first Floquet sidebands $(m=0,\pm1)$ contribute significantly to the dynamics.

In the interaction picture, the reduced density matrix evolves according to

\begin{align}
\frac{d}{dt}\rho(t)=\sum_{k=h,c} \pazocal{L}_k^{\rm F}\left[\rho(t)\right],,
\end{align}
with 
\begin{align}\label{eq:Lindblad dissipatiors}
 \pazocal{L}_k^{\rm F}\left[\rho(t)\right]=  \frac{\Gamma_k^+}{2}\Bigg( \left[ \sigma^+_k, \rho(t) \sigma^-_k\right]+ \left[ \sigma^+_k\rho(t), \sigma^-_k\right]\Bigg)+ \frac{\Gamma_k^-}{2} \Bigg( \left[ \sigma^-_k, \rho(t) \sigma^+_k\right]+ \left[ \sigma^-_k\rho(t), \sigma^+_k\right]\Bigg)
\end{align}
denotes the Floquet dissipator associated with bath (k). The effective excitation and relaxation rates are given by
\begin{equation}
    \Gamma_k^\pm = \sum_{m = -\infty}^{+\infty} P(m) \gamma_k^\pm (\omega_0 + m \omega_{\rm F})
\end{equation}
where $\gamma_k^\pm(\omega)$ are the bath-induced excitation (+) and relaxation (-) rates at frequency $\omega$, which satisfy the KMS condition, while $P_m$ denotes the weight associated with the Floquet mth sideband.

As illustrated in Fig.~\ref{fig:sub2}, the bath spectra are assumed to be non-overlapping and centered around the frequencies 
\begin{align}
\omega_h=\omega_0+\omega_{\rm F},  \quad  \omega_c=\omega_0-\omega_{\rm F}, \,\,\, \text{with}\quad\omega_c >0,
\end{align}
such that $\gamma_k^\pm(\omega_0)=0$. Consequently, only the sidebands $m=\pm1$ contribute, with corresponding probabilities

\begin{align}
P_{\pm}=\abs{\varepsilon_{\pm}}^2, \quad \varepsilon_{\pm} = \frac{1}{\tau} \int_0^\tau e^{i \Omega \int_0^t \sin(\omega_{\rm F} s) ds} e^{\pm i \omega_{\rm F} t}
\end{align}

These quantities correspond to the probabilities of shifting the bath coupling spectra $\gamma_{h(c)}^\pm$ by $\pm\omega_{\rm F}$ with respect to the average frequency $\omega_0$. The amplitudes $\varepsilon_\pm$ can be expressed in terms of Bessel functions as
\begin{equation}
    \varepsilon_\pm = \frac{1}{\tau} \int_0^\tau dt \ \exp[i \left(\frac{\Omega}{\omega_{\rm F}} (1-\cos(\omega_{\rm F} t)) \pm \omega_{\rm F} t\right)] = -ie^{ i \Omega/ \omega_{\rm F}}  J_1\left(\frac{\Omega}{\omega_{\rm F}}\right),
\end{equation}
and therefore $P_\pm =  J^2_1\left(\frac{\Omega}{\omega_{\rm F}}\right)$.

\subsubsection{Heat currents $\&$ Power}
It is sufficient to adopt the approach proposed in \cite{PhysRevE.87.012140}, in which one can find some derived expressions for the heat currents $J_{\rm Floquet}^h$ and $J_{\rm Floquet}^c$ as well as for the power $P_{\rm Floquet}$. Note that these quantities are found in the steady-state regime. Adapting their results with our notation, we have
\begin{align}
J_{\rm Floquet}^h&= \omega_h P_+\gamma_h^-\frac{e^{-\beta_h \omega_h}-r_{ss}}{r_{ss}+1} \\
J_{\rm Floquet}^c&= \omega_c P_-\gamma_c^-\frac{e^{-\beta_c \omega_c}-r_{ss}}{r_{ss}+1}
\end{align}
$r_{ss}$ is called the population ratio and has the form
\begin{align}
r_{ss}= \frac{P_+ \gamma_h^+ +P_- \gamma_c^+}{P_+ \gamma_h^- +P_- \gamma_c^-}
\end{align}
Plugging this expression into $J_{\rm Floquet}^h$ and $J_{\rm Floquet}^c$, we obtain
\begin{equation}
    J_{\rm Floquet}^h = \omega_h \langle \dot {\tilde{p}} \rangle_{ss}, \quad  J_{\rm Floquet}^c = -\omega_c \langle \dot {\tilde{p}} \rangle_{ss}\,,
\end{equation}
where
\begin{equation}
    \langle \dot {\tilde{p}} \rangle_{ss} = \frac{P_- P_+  \gamma_c^-  \gamma_h^- (e^{-\beta_h \omega_h}-e^{-\beta_c\omega_c})}{(1+e^{-\beta_c \omega_c}) P_-  \gamma_c^- + (1+e^{-\beta_h \omega_h}) P_+  \gamma_h^-}\,.
\end{equation}
The corresponding power is given by
\begin{align}
P_{\rm Floquet}=\frac{\left( \omega_h-\omega_c\right) P_- P_+  \gamma_c^- \gamma_h^- (e^{-\beta_h \omega_h}-e^{-\beta_c \omega_c})}{(1+e^{-\beta_c \omega_c}) P_- \gamma_c^- +(1+e^{-\beta_h \omega_h}) P_+  \gamma_h^-},\quad \text{with}\,\,\,\beta_h\omega_h<\beta_c\omega_c.
\end{align}
After simplification, the extracted power defined above reads as follows
\begin{align}
P_{\rm Floquet}=\frac{ J^2_1\left(\frac{g}{\omega_{\rm F}}\right)}{\tau_h + \tau_c}W_{\rm discrete}.
\end{align}

\subsubsection{Operational time}

\begin{figure}[t!]
\centering
\includegraphics[width=.8\linewidth]{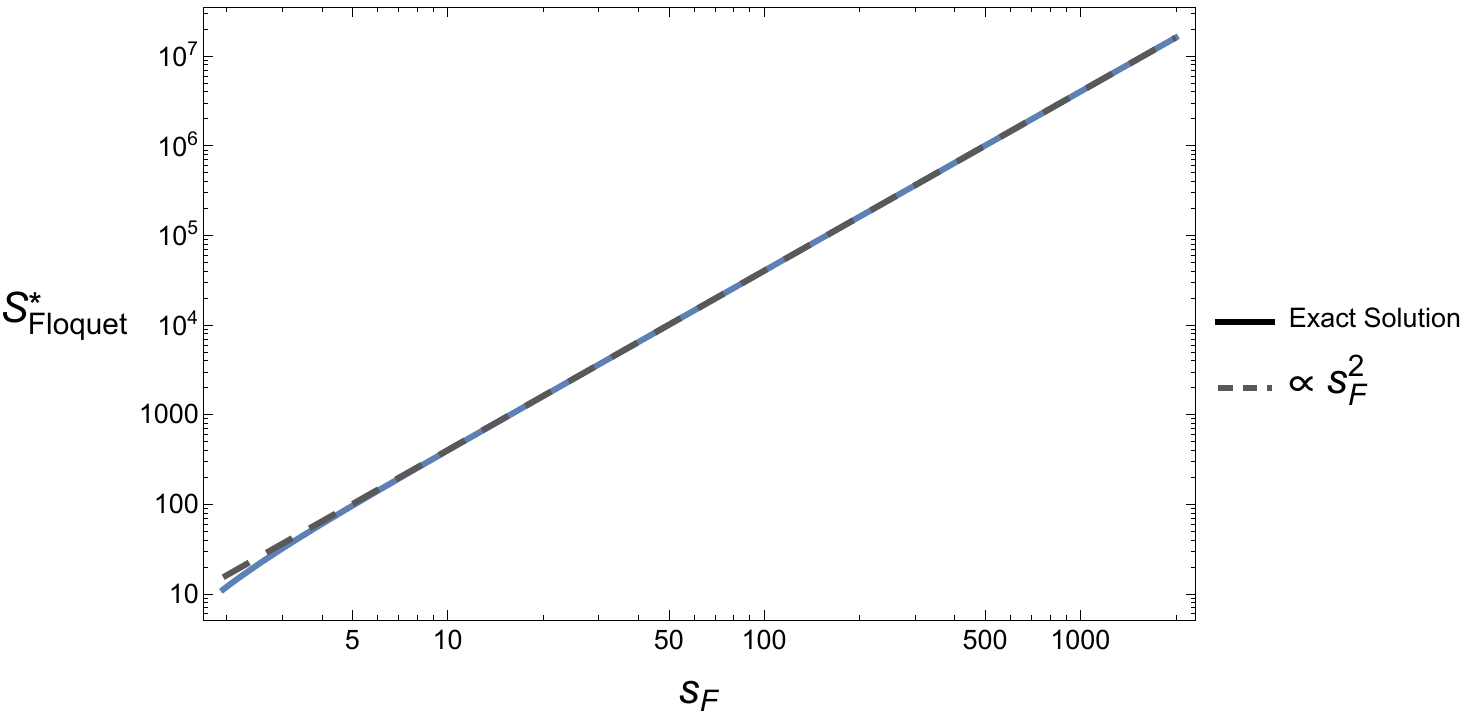}
\caption{
Log-log plot of the dimensionless operational timescale
\(S_{\rm Floquet}^*\) as a function of the dimensionless parameter
\(s_{\rm F}\).
The solid curve represents the numerical result, while the dashed curve corresponds to the asymptotic scaling
\(S_{\rm Floquet}^*-1 \propto s_F^2\).
The logarithmic scales emphasize the asymptotic power-law behavior.
}
\label{fig:asymptotic-floquet}
\end{figure}

The corresponding characteristic time of the Floquet engine is
\begin{align}
T^*_{\rm Floquet}=\frac{W_{\rm discrete}}{P_{\rm Floquet}}
=\frac{\tau_h +\tau_c}{J^2_1\left(\frac{\Omega}{\omega_{\rm F}}\right)},
\end{align}
where \(\tau_h\) and \(\tau_c\) denote the thermalization timescales defined in Eq.~\eqref{eq: thermalization timescales}. This expression can be recast in dimensionless form by introducing the parameter $s_F=\omega_{\rm F}/\Omega$, yielding
\begin{align}
S^*_{\rm Floquet}(s_{\rm F})= \frac{T^*_{\rm Floquet}}{\tau_h + \tau_c}=J_1^{-2}\left(s_F^{-1}\right).
\end{align}
The asymptotic behavior of \(S^{*}_{\rm Floquet}\) follows from the large-argument expansion of the Bessel function. In particular, for \(s_{\rm F}\gg1\),
\begin{align}
S^{*}_{\rm Floquet}(s_{\rm F})= 1+4s_{\rm F}^2 +\pazocal{O}(s^{-2}_{\rm F}).
\end{align}
To verify this scaling, Fig.~\ref{fig:asymptotic-floquet} presents a log-log representation of the dimensionless operational timescale together with its asymptotic approximation. In this representation, the quadratic dependence appears as a straight line with slope \(2\), in excellent agreement with the numerical results.

\subsection{Always on engine (transverse coupling)}\label{model: always-on}
We now consider an always-on engine whose working medium consists of two weakly interacting qubits. The hot and cold qubits have transition frequencies $\omega_h$ and $\omega_c$ and remain continuously coupled to reservoirs at inverse temperatures $\beta_h$ and $\beta_c$, respectively. A periodic modulation resonantly activates excitation exchange between the qubits. Consequently, coherent work conversion and thermalization occur simultaneously.

The dynamics of the system are described by the time-dependent Hamiltonian
\begin{align}\label{eq:driven Hamiltonian}
{H}_{\rm tran}(t)={H}_{0}+ {V}(t)\,,
\end{align}
with
\begin{align}
{H}_{0} &= {H}_h + {H}_c = \omega_h \sigma^+_h \sigma^-_h +  \omega_c \sigma^+_c \sigma^-_c \\
{V}(t) &= \frac{\Omega}{2} \left(\sigma^-_h \sigma^+_c e^{ i \omega_d t} + h.c. \right) \label{eq:drive}.
\end{align}
Here, $\sigma_{h,c}^{\pm}$ denote the ladder operators of the hot and cold qubits, respectively, $\Omega$ is the driving strength, and $\omega_d=\omega_h-\omega_c$ is the resonant driving frequency. Throughout this section, we assume $\Omega \ll \omega_h, \omega_c$, together with the weak system-bath coupling and Markovian assumptions required by the local master equation.

%\textcolor{red}{for four-strokes case, going uphill means ramping up and going downhill is either ramping down or tapering off...}
\subsubsection{Master equation}

The dynamics of the two-qubit system are described by the local master equation, which in the interaction picture reads
\begin{align}
\dot{\rho_I}(t) =-i\left[{V}(0),\,\rho_I(t) \right]+ \pazocal{L}_h\left[ \rho_I(t)\right]+\pazocal{L}_c\left[ \rho_I(t)\right]\,,
\end{align}
where 
\begin{align}\label{eq: dissipator two qubits}
\pazocal{L}_k\left[ \rho_I(t)\right]= \gamma^+_k\left(\sigma^+_k \rho_I(t)\sigma^-_k -\frac{1}{2}\left\{\sigma^-_k\sigma^+_k,\rho_I(t) \right\}\right)+\gamma^-_k\left(\sigma^-_k \rho_I(t)\sigma^+_k -\frac{1}{2}\left\{\sigma^+_k\sigma^-_k,\rho_I(t) \right\}\right)\,.
\end{align}
$k=h,c$, $\gamma^-_k\left( \gamma^+_k \right)$ $k-$bath damping (pumping) rates that satisfy the detailed balance relation $\gamma^+_k/\gamma^-_k=e^{-\beta_k \omega_k}$.

Equivalently, the local master equation governing the evolution of an arbitrary observable $\pazocal{O}_I$ in the interaction picture follows from the Heisenberg equation of motion
\begin{equation}
\frac{d}{dt}\pazocal{O}_I=i\left[{V}(0),\,\pazocal{O}_I\right]+\pazocal{L}_h^{\dagger}\left[ \pazocal{O}_I\right]+\pazocal{L}_c^{\dagger}\left[ \pazocal{O}_I\right]:=\pazocal{L}^{\dagger}\left[ \pazocal{O}_I\right]\,,
\end{equation}
with
\begin{align}
\pazocal{L}_k^{\dagger}\left[ \pazocal{O}_I\right]= \gamma^+_k\left(\sigma^-_k \pazocal{O}_I\sigma^+_k -\frac{1}{2}\left\{\sigma^-_k\sigma^+_k,\pazocal{O}_I \right\}\right)+\gamma^-_k\left(\sigma^+_k \pazocal{O}_I\sigma^-_k -\frac{1}{2}\left\{\sigma^+_k\sigma^-_k,\pazocal{O}_I \right\}\right)\,,
\end{align}

\subsubsection{Heat currents $\&$ Power}\label{eq: steady-state prob. flow}

The first law is expressed by the equation:
\begin{align}
\dot{E}(t)=P(t)+J_h(t)+J_c(t)
\end{align}
where the internal energy $E(t)$, power $P(t)$, and heat currents $J_k(t)$ are given by
\begin{align}
E(t)&=\Tr\left[ {H}(t) \rho_I(t)\right]\,, \\
P(t)&=\Tr\left[\frac{d}{dt}{V}(t)\big|_{t=0}\rho_I(t) \right],\, \quad \quad 
J_h(t)=\Tr \left[\pazocal{D}^{\dagger}_h\left[{H}_0\right] \rho_I(t)\right],\quad  \quad J_c(t)=\Tr \left[\pazocal{D}^{\dagger}_c\left[{H}_0\right] \rho_I(t)\right]\label{eq8}
\end{align}
In the steady state regime, Eq. \eqref{eq: dissipator two qubits} becomes $\pazocal{L}\left[\rho_I^{ss} \right]=0$. Therefore, for any observable $\pazocal{O}_I$, $\Tr\left[\pazocal{L}^{\dagger}\left[ \pazocal{O}_I \right]\rho_I^{ss} \right]=0$. Let us introduce the
operator 
\begin{equation}
\dot{{p}}= i\frac{\Omega}{2}\left(\sigma^+_h\sigma^-_c-\sigma^-_h\sigma^+_c\right)\,.
\end{equation}
Defining $p_{ij}=\bra{i_h j_c}\rho_I^{ss}\ket{i_h j_c}$, the condition $\Tr\left[\pazocal{L}^{\dagger}\left[ \dot{{p}} \right]\rho_I^{ss} \right]=0$ leads to :
\begin{equation}
\Tr\left[\pazocal{L}^{\dagger}\left[ \dot{{p}} \right]\rho_I^{ss} \right]= \Omega^2\left(p_{10}-p_{01} \right) -\left( \Gamma_h +\Gamma_c\right)\langle\dot{{p}}\rangle_{ss}=0
\end{equation}
where $\Gamma_k=\gamma^+_{k}+\gamma^-_k$ and we use the short-hand notation$\langle\dot{{p}}\rangle_{ss}=\Tr\left[ \dot{{p}} \rho_I^{ss}  \right]$ which represents the stationary probability current. The condition $\Tr\left[\pazocal{L}^{\dagger}\left[ \pazocal{O}_I \right]\rho_I^{ss} \right]=0$ applied to projectors $\ket{i}\bra{j}$ for $i,j \in \{0,1\}$ translates to the following set of equations 
\begin{align}
-\left(\gamma^+_h+\gamma^+_c\right)p_{00}+ \gamma^-_h p_{10} + \gamma^-_c p_{01}&=0, \\
-\left(\gamma^+_h+\gamma^-_c\right)p_{01}+\gamma^-_h p_{11} + \gamma^+_c p_{00} +\langle\dot{p}\rangle_{ss}&=0, \\
-\left(\gamma^-_h+\gamma^+_c\right)p_{10}+\gamma^-_c p_{11} + \gamma^+_h p_{00} -\langle\dot{p}\rangle_{ss}&=0 ,\\
-\left(\gamma^-_h+\gamma^-_c\right)p_{00}+ \gamma^+_h p_{01} + \gamma^+_c p_{10}&=0 ,
\end{align}
which, upon solving, yields
\begin{align}
  \langle\dot{{p}}\rangle_{ss}=\frac{ \left(\gamma^-_c \Gamma_h - \gamma^-_h \Gamma_c \right)}{\left( \Gamma_h+\Gamma_c\right)\left( 1+\frac{ \Gamma_h \Gamma_c}{\Omega^2} \right)}  
\end{align}
This expression can be rewritten as
\begin{align}\label{eq18}
\langle\dot{p}\rangle_{ss}= \frac{\Gamma_h\Gamma_c}{\left(\Gamma_h+\Gamma_c \right)}\frac{\delta p}{\left( 1+\frac{ 2\Gamma_h \Gamma_c}{\Omega^2} \right)}
\end{align}
with $\delta p=\frac{a_h-a_c}{\left(1+a_h\right)\left(1+a_c\right)}$ and $a_k=e^{-\beta_k \omega_k}\,\,(k \in \{h,c\})$. Heat currents in the steady state regime can be obtained using \eqref{eq18} as follows
\begin{alignat}{2}
\displaystyle
J_h&=\omega_h \langle\dot{{p}}\rangle_{ss}&=\omega_h  \frac{\Gamma_h\Gamma_c}{\left(\Gamma_h+\Gamma_c \right)}\frac{\delta p}{\left( 1+\frac{ \Gamma_h \Gamma_c}{\Omega^2} \right)} \\
J_c&=-\omega_c \langle\dot{{p}}\rangle_{ss}&=-\omega_c  \frac{\Gamma_h\Gamma_c}{\left(\Gamma_h+\Gamma_c \right)}\frac{\delta p}{\left( 1+\frac{ \Gamma_h \Gamma_c}{\Omega^2} \right)}
\end{alignat}
In the same regime, power is obtained using $P=\left(J_h+J_c\right)$ and reads
\begin{align}
P=\left(\omega_h-\omega_c\right)  \frac{\Gamma_h\Gamma_c}{\left(\Gamma_h+\Gamma_c \right)}\frac{\delta p}{\left( 1+\frac{ \Gamma_h \Gamma_c}{\Omega^2} \right)}
\end{align}
The efficiency of our engine yields
\begin{align}
\eta=\frac{P}{J_h}=1-\frac{\omega_c}{\omega_h}
\end{align}
\subsubsection{Operational time}
In the steady-state regime, the corresponding characteristic operational time of the always-on engine is
\begin{align}
T^*_{\rm always-on}&= \frac{W_{\rm discrete}}{P_{\rm always-on}}=\big(\tau_h+\tau_c \big)\left(1+\frac{t_{\pi}^2}{\pi^2 \tau_h \tau_c} \right),
\end{align} 
where \(t_\pi=\pi/\Omega\) denotes the duration of the \(\pi\)-pulse. Introducing the dimensionless parameter
\[s_\pi=\frac{t_\pi}{\sqrt{\tau_h\tau_c}},\] the corresponding dimensionless operational time becomes
\begin{align}
S^*_{\rm always-on}(s_\pi)= \frac{T^*_{\rm always-on}}{\tau_h + \tau_c} = 1+\frac{s_\pi^2}{\pi^2}.
\end{align}
Unlike previous models, no optimization is required in this case, as the dimensionless operating time is given analytically and naturally scales quadratically with \(s_\pi\) as shown in Fig \ref{always-on} below.

\begin{figure}[h!]
\centering
\includegraphics[width=0.8\linewidth]{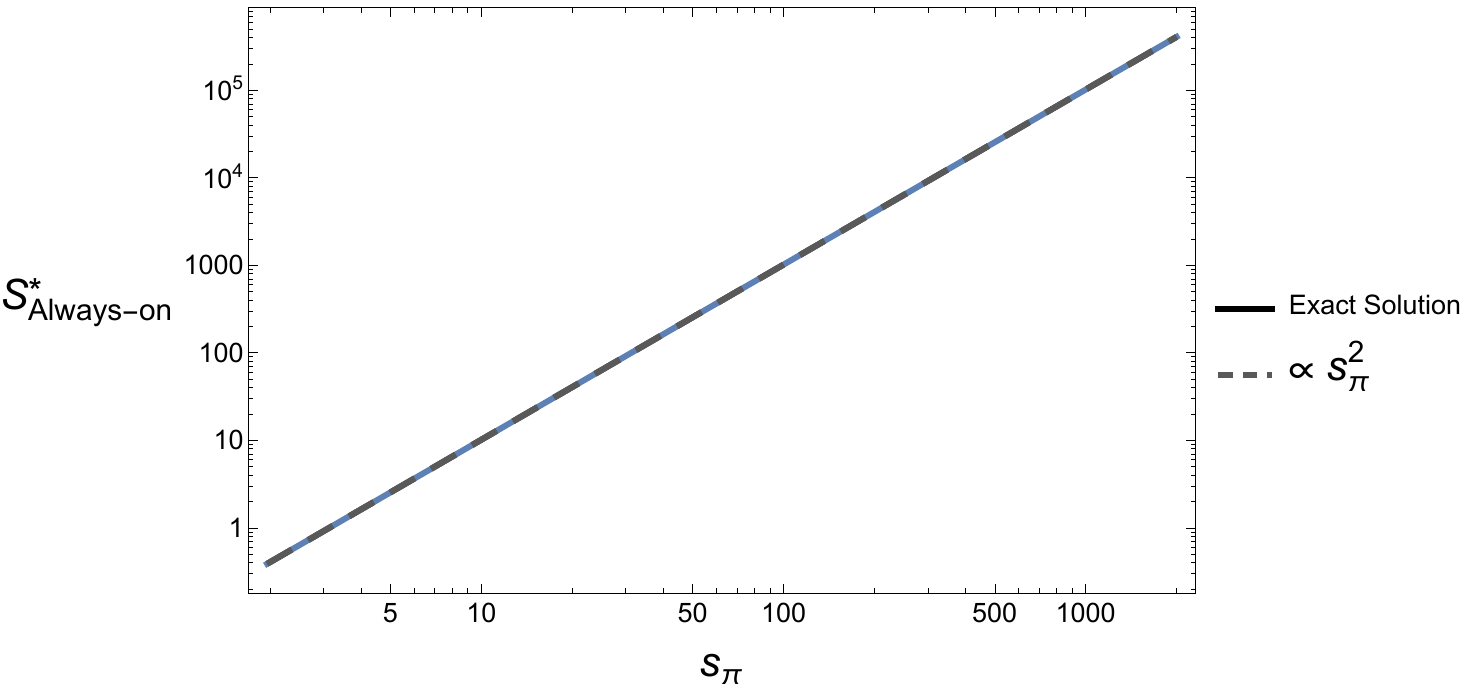}
\caption{Log-log plot of the dimensionless operational timescale $S_{\mathrm{always\text{-}on}}^{*}$ as a function of the dimensionless parameter $s_{\pi}$. The solid curve represents the exact result, while the dashed curve corresponds to the asymptotic scaling $S_{\mathrm{always\text{-}on}}^{*}-1\propto s_{\pi}^{2}$. The logarithmic scales emphasize the quadratic power-law behavior in the large- and low-$s_{\pi}$ regime.}
\label{always-on}
\end{figure}

\subsection{On-and-off engine (transverse coupling)}\label{model: on-and-off}
In this section, we provide additional details regarding the implementation of the on-off driving protocol discussed in Section~\ref{On-and-off model} of the main text.

The protocol consists of two alternating stages: a coherent driving stroke and a thermalization stroke. During the coherent stage, the system is isolated from both thermal reservoirs and evolves solely under the action of the driving Hamiltonian given in Eq.~\eqref{eq:driven Hamiltonian}. In contrast, during the thermalization stroke, the external drive is switched off and each qubit interacts only with its respective thermal reservoir, thereby relaxing towards its local thermal equilibrium state.

Both stages are assumed to occur over finite time intervals, indicated by $t_\pi$ for the coherent driving stroke and $t_{hc}$ for the thermalization stroke as defined in the main text. Together, these stages form a cycle with total duration $\tau_{\rm on-off}=t_{\pi}+t_{hc}$.

\subsubsection{Stage 1: Coherent drive}
In this stage, each qubit is initially prepared in a local thermal state, such that the two-qubit system is initially described by

\begin{align}
\rho_0^{(1)}=\frac{e^{-\beta_h {H}_h}}{\Tr\Big[e^{-\beta_h {H}_h}\Big]} \otimes \frac{e^{-\beta_c {H}_c}}{\Tr\Big[e^{-\beta_c {H}_c}\Big]}. 
\end{align}
Since the system is decoupled from both reservoirs during the coherent stroke, its evolution is governed solely by the Hamiltonian dynamics
\begin{align}
\frac{d \rho(t)}{dt}=-i [{H}_{\rm tran}(t), \rho(t)]\,.
\end{align}
Moving to the interaction picture and subsequently returning to the Schrödinger picture, the solution can be written as
\begin{align}\label{eq: work stroke state}
\rho(t)={U}_0(t){U}_I(t)\rho_0^{(1)}{U}_I(t)^{\dagger}{U}_0(t)^{\dagger}\,,
\end{align}
where 
\begin{align}
{U}_0(t) &=\exp\left[-i {H}_0 t\right],\hspace{0.6cm}
{U}_I(t)=\exp[-i {H}_I t], \\
\text{and}\,\,\,\, {H}_I&={V}(0)\,.
\end{align}
This protocol lasts for the duration $t_\pi=\pi/\Omega$, thus implementing a complete population transfer between the states $\ket{01}$ and $\ket{10}$. The state at the end of the stroke is therefore
\begin{align}
\rho_1^{(1)}= {U}_{t_\pi} \rho_0^{(1)} {U}_{t_\pi}^{\dagger}= {U}_0(t_\pi){U}_I(t_\pi) \rho_0^{(1)}{U}_I(t_\pi)^{\dagger}{U}_0(t_\pi)^{\dagger}
\end{align}
where
\begin{align}
{U}_{t_\pi}
={U}_0(t_\pi){U}_I(t_\pi).
\end{align}
The average work exchanged during this stage is given by
\begin{align}
    W_{\rm on-off}^{(1)} &= -\Tr[{H}_0 (\rho_1^{(1)} - \rho_0^{(1)})] =\big(\omega_h-\omega_c\big) \delta p
\end{align}
with $\delta p$ given in Eq.~\eqref{eq:deltap} in the main text.

%{Q^{h,c \ (m)}_{\rm on-off}} &= \Tr[H_{h,c} (\rho_2^{(m)} - \rho_1^{(m)})]

\subsubsection{Stage 2: Thermalization stage}
Following the coherent stroke, the external drive is switched off, and each qubit is reconnected to its respective thermal reservoir. During this stage, which lasts for a duration $t_{hc}$, the dynamics of the two-qubit system is described by the Markovian master equation
\begin{align}
\frac{d}{dt}\rho(t) =-i [{H}_0, \rho(t)]+ \pazocal{L}_h[\rho(t)]+\pazocal{L}_c[\rho(t)],\,
\end{align}
where the dissipators $\pazocal{L}_k[\cdot]$ are given in \eqref{eq: dissipator two qubits}. Restricting the analysis to the population dynamics, the master equation reduces to the following set of coupled linear differential equations
\begin{align}\label{eq:populations}
\begin{bmatrix}
\dot{p}^{(1)}_{00}(t) \\ \dot{p}^{(1)}_{01}(t) \\ \dot{p}^{(1)}_{10}(t) \\ \dot{p}^{(1)}_{11}(t)
\end{bmatrix}
=
\begin{pmatrix}
-(\gamma^{+}_{h} + \gamma^{+}_{c}) &  \gamma^{-}_{c} & \gamma^{-}_{h} & 0\\ \gamma^{+}_{c} & -(\gamma^{+}_{h} + \gamma^{-}_{c}) & 0 & \gamma^{-}_{h}\\ \gamma^{+}_{h} & 0 & -(\gamma^{+}_{c} + \gamma^{-}_{h}) & \gamma^{-}_{c} \\ 0 & \gamma^{+}_{h} & \gamma^{+}_{c} & -(\gamma^{-}_{h} + \gamma^{-}_{c})
\end{pmatrix}
\begin{bmatrix}
p^{(1)}_{00}(t) \\ p^{(1)}_{01}(t) \\ p^{(1)}_{10}(t) \\ p^{(1)}_{11}(t)
\end{bmatrix}
\end{align}
The thermalization stroke starts from the state produced by the preceding coherent evolution,
 \begin{align}
\rho(t_\pi)= \rho_1^{(1)}
\end{align}
with 
\begin{align}
\rho_1^{(1)}=\text{diag}\left( p_{00}^{(0)}, p_{10}^{(0)}, p_{01}^{(0)}, p_{11}^{(0)} \right)
\end{align}
and 
\begin{align}
p_{00}^{(0)}=\frac{1}{\left(1+a_c\right)\left(1+a_h\right)}, \quad p_{01}^{(0)}=\frac{a_c}{\left(1+a_c\right)\left(1+a_h\right)}, \quad p_{10}^{(0)}=\frac{a_h}{\left(1+a_c\right) \left(1+a_h\right)}, \quad p_{11}^{(0)}=\frac{a_c a_h}{\left(1+a_c\right)\left(1+a_h\right)}
\end{align}
$a_k=e^{-\beta_k \omega_k}$ $(k=h,c)$.

Let us define a vector $\mathbf{p}(t)=\left(p_{00}^{(1)}(t),\, p_{01}^{(1)}(t), \, p_{10}^{(1)}(t), \, p_{11}^{(1)}(t)\right)$ as a solution of \eqref{eq:populations} at time $t$. At the end of the second stage, the solution has the form
\begin{align}\label{eq: heat stroke population}
\mathbf{p}_2^{(1)} = e^{M t_{hc}} \mathbf{p}_1^{(1)}\,,
\end{align}
where 
\begin{align*}
M= 
\begin{pmatrix}
-(\gamma^{+}_{h} + \gamma^{+}_{c}) &  \gamma^{-}_{c} & \gamma^{-}_{h} & 0\\ \gamma^{+}_{c} & -(\gamma^{+}_{h} + \gamma^{-}_{c}) & 0 & \gamma^{-}_{h}\\ \gamma^{+}_{h} & 0 & -(\gamma^{+}_{c} + \gamma^{-}_{h}) & \gamma^{-}_{c} \\ 0 & \gamma^{+}_{h} & \gamma^{+}_{c} & -(\gamma^{-}_{h} + \gamma^{-}_{c})
\end{pmatrix}
\end{align*}
M can be decomposed as $RDR^{-1}$, with \\ 
\begin{align*}
R&= 
\begin{pmatrix}
\frac{\gamma^{-}_{c} \gamma^{-}_{h}}{\gamma^{+}_{c}\gamma^{+}_{h}} & -\frac{\gamma^{-}_{h}}{\gamma^{+}_{h}} & -\frac{\gamma^{-}_{c}}{\gamma^{+}_{c}} & 1 \\\\
\frac{ \gamma^{-}_{h}}{\gamma^{+}_{h}} & \frac{\gamma^{-}_{h}}{\gamma^{+}_{h}} & -1 & -1 \\\\
\frac{ \gamma^{-}_{c}}{\gamma^{+}_{c}} & -1 & \frac{ \gamma^{-}_{c}}  {\gamma^{+}_{c}} & -1  \\\\
1 & 1 & 1 & 1
\end{pmatrix}
 \hspace{1cm}
\text{and} 
\hspace{0.7cm}
D&= 
\begin{pmatrix}
0 & 0 & 0 & 0\\\\
0 & -\gamma^{-}_{c} -\gamma^{+}_{c} & 0 & 0 \\\\
0 & 0 & -\gamma^{-}_{h}-\gamma^{+}_{h} & 0 \\\\
0 & 0 & 0 & -\gamma^{-}_{c} -\gamma^{-}_{h}-\gamma^{+}_{c} - \gamma^{+}_{h}
\end{pmatrix}
\end{align*}
\\\\
so that 
\begin{align}\label{eq: populations}
\mathbf{p}_2^{(1)} = R e^{D t_{hc}}  R^{-1}\mathbf{p}_1^{(1)}\,.
\end{align}
Populations are the only relevant elements of the density matrix to evaluate heat, which is expressed as follows
\begin{align}
{Q^{h,c \ (1)}_{\rm on-off}} &= \Tr[\hat{H}_{h,c} (\rho_2^{(1)} - \rho_1^{(1)})] 
\end{align}
and result as

\begin{align}
Q^{(1)}_h&=\omega_h\delta p \left(1-e^{-\Gamma_h t_{hc}}\right) \\
Q^{(1)}_c&=-\omega_c\delta p \left(1-e^{-\Gamma_c t_{hc}}\right) .
\end{align}

%\textcolor{red}{add explicit forms}.
\subsubsection{Asymptotic limit}
We are now interested in the repeated operation of the engine over many cycles. Denoting by $\rho_i^{(m)}$ the state of the system at stage $i$ of the m-th cycle, the sequence of states generated by the protocol can be represented schematically as follows
\begin{equation}
\underbrace{
\rho_0^{(1)}
\xrightarrow[\text{stroke}]{\text{work}}
\rho_1^{(1)}
\xrightarrow[\text{stroke}]{\text{heat}}
\rho_2^{(1)}=\rho_0^{(2)}
}_{\text{1st cycle}}
\underbrace{
\xrightarrow[\text{stroke}]{\text{work}}
\rho_1^{(2)}
\xrightarrow[\text{stroke}]{\text{heat}}
\rho_2^{(2)}=\rho_0^{(3)}
}_{\text{2nd cycle}}
\underbrace{\xrightarrow[\text{stroke}]{\text{work}}}_{\dots}
\underbrace{\dots
\xrightarrow[\text{stroke}]{\text{heat}}
\rho_2^{(m)}}_{\text{nth cycle}}
\end{equation}
The evolution associated with a single cycle was discussed in Eq.~\eqref{On-off sequence}. Using \eqref{eq: work stroke state} and \eqref{eq: heat stroke population}, we can construct the map describing the evolution over an arbitrary number of cycles.

Since the density matrix remains diagonal throughout the evolution, it is sufficient to focus only on the population dynamics. To this end, we first consider the work stroke of the first cycle and introduce the operator ${S}$, defined through
\begin{align}
\rho_1^{(1)}={S}\rho_0^{(1)}{S}^{\dagger}
\end{align}
This operator has the following matrix form on the basis $\{\ket{00},\ket{01},\ket{10},\ket{11}\}$
\begin{align}
{S}=
\begin{pmatrix}
1 & 0 & 0 & 0 \\
0 & 0 & 1 & 0\\
0 & 1 & 0 & 0 \\
0 & 0 & 0 & 1
\end{pmatrix}
\end{align}
Populations are obtained using projective measurements:
\begin{align}
\left(\mathbf{p}_1^{(1)}\right)_{ij}=\Tr\left[\ket{i}\bra{j}{S}\rho_0^{(1)}{S}^{\dagger}\right]\quad (i,j \in \{0,1\})
\end{align}
Together with \eqref{eq:populations}, our new map yields
\begin{align}
\Lambda: t_{hc} \to R e^{D t_{hc}} R^{-1} S\,.
\end{align}
For the $m^{th}$ cycle, we repeatedly apply this map to the initial state m times. When m becomes large, the system reaches an asymptotic state 
\begin{align}\label{on-off asymptotic}
\displaystyle 
\lim_{n \to \infty} {[\Lambda]}^n(\mathbf{p}_0^{(1)})&=
\begin{pmatrix}
p_{00}^{asy} & p_{01}^{asy} & p_{10}^{asy} & p_{11}^{asy} \\\\
p_{00}^{asy} & p_{01}^{asy} & p_{10}^{asy} & p_{11}^{asy} \\\\
p_{00}^{asy} & p_{01}^{asy} & p_{10}^{asy} & p_{11}^{asy} \\\\
p_{00}^{asy} & p_{01}^{asy} & p_{10}^{asy} & p_{11}^{asy}
\end{pmatrix}
\end{align}
with 
\begin{align}
p_{00}^{asy}=& \frac{f_+(t_{hc{}})g_{-}(t_{hc})}{\left[h(t_{hc})\right]^2}, \quad
p_{01}^{asy}= \frac{f_+(t_{hc})g_{+}(t_{hc})}{\left[h(t_{hc})\right]^2}, \quad
p_{10}^{asy}= \frac{f_-(t_{hc})g_{-}(t_{hc})}{\left[h(t_{hc})\right]^2}, \quad 
p_{11}^{asy}=\frac{f_-(t_{hc})g_{+}(t_{hc})}{\left[h(t_{hc})\right]^2}
\end{align}
where
\begin{align}
f_{\pm}(t_{hc})&= -\gamma^{\mp}_c\Gamma_h+e^{t_{hc}\left(\Gamma_h+\Gamma_c\right)}\gamma^{\mp}_h\Gamma_c \pm e^{t_{hc}\Gamma_c}\gamma^-_c\gamma^-_h\left(a_h-a_c\right) \\
h(t_{hc})&=\left[-1+e^{t_{hc}\left(\Gamma_h+\Gamma_c\right)}\right]\Gamma_h \Gamma_c
\end{align}
It is worth mentioning that $ p_{00}^{asy}+ p_{01}^{asy} + p_{10}^{asy} + p_{11}^{asy}=1$. From Eq. \eqref{on-off asymptotic}, the state of the system in the asymptotic limit reads:
\begin{align}
\rho_{asy} = 
\begin{pmatrix}
p_{00}^{asy} & 0 & 0 & 0 \\
0 & p_{01}^{asy} & 0 & 0 \\
0 & 0 & p_{10}^{asy} & 0 \\
0 & 0 & 0 & p_{11}^{asy}
\end{pmatrix}
\end{align}
This state can be assumed to be the initial state of a new cycle of our engine. Thus, the work exchanged during the first stage of this new cycle is obtained as follows:

\begin{align}
 W_{\rm on-off}^{asy}  &= (\omega_h - \omega_c)\Big(p_{10}^{asy}- p_{01}^{asy}\Big) \\
&=\frac{(\omega_h - \omega_c)\Bigg( -1 + e^{t_{hc}\Gamma_c} \Bigg) \Bigg( -1 + e^{t_{hc}\Gamma_h} \Bigg) }{ \Bigg(-1 + e^{t_{hc}\left(\Gamma_h +\Gamma_c \right)} \Bigg) }\delta p=(\omega_h-\omega_c)\delta p_{asy}.
\end{align}

In this limit, the power is given by: 
\begin{align}
   P_{\rm on-off} &= \frac{W_{\rm on-off}^{asy}}{\tau_{\rm on-off}} = \frac{\left(1 - e^{ -t_{hc} /\tau_h} \right) \left(1 + e^{ - t_{hc}/\tau_c} \right)}{(t_\pi + t_{hc}) (1 - e^{-t_{hc}\left(1/\tau_h +1/\tau_c \right)} )} W_{\rm discrete}
\end{align}
\subsubsection{Operational time}
In the asymptotic regime,  the characteristic operational time of this engine is given by
\begin{align}
T^*_{\rm on-off}&= \frac{W_{\rm discrete}}{P_{\rm on-off}} = \frac{(t_\pi+t_{hc})(1 - e^{-t_{hc}\left(1/\tau_h +1/\tau_c \right)})}{\left(1 - e^{ -t_{hc} /\tau_h} \right) \left(1 - e^{ - t_{hc}/\tau_c} \right)}
\end{align}
Introducing the dimensionless thermalization time $s_{hc} =t_{hc}/\sqrt{\tau_h\tau_c}$ and the dimensionless modulation timescale $s_\pi$ defined in the previous subsection, the corresponding dimensionless operational time becomes
\begin{align}
\frac{T^*_{\rm on-off}}{\tau_h + \tau_c} =\frac{\kappa}{2} \frac{( s_\pi+s_{hc}) \left(1 - e^{ -2 s_{hc}/\kappa} \right)}{\left(1 - e^{ -\sigma s_{hc}} \right)\left(1 - e^{ -s_{hc} / \sigma} \right)},
\end{align}
where \(\sigma=\kappa^{-1}+\sqrt{\kappa^{-2}-1}.\) The optimal dimensionless operational time is then obtained by minimizing over the thermalization variable \(s_{hc}\) for fixed values of \(s_\pi\), namely,
\begin{align}
S^*_{\rm on-off}(s_\pi)=\min_{s_{hc}\ge0}\left(\frac{T^*_{\rm on-off}}{\tau_h+\tau_c}\right).
\end{align}
This expression has the same mathematical structure as Eq.~\eqref{eq:recast-four-stroke time}, and therefore a closed analytical expression cannot be obtained for the optimized operational time. Nevertheless, useful analytical information can be extracted in the limit of short thermalization times. Expanding the exponentials to the first order in \(s_{hc}\), we obtain
\begin{align}
S^*_{\rm on-off}(s_\pi) \approx \frac{\kappa}{2}\frac{\left(s_\pi + s_{hc}\right)2 s_{hc}}{\kappa s_{hc}^2} \geq 1
\end{align}

Hence, the optimized operational time is bounded from below by
\begin{align}
S^*_{\rm on-off}(s_\pi) \ge S^{*}_{\rm on-off}(0)= 1.
\end{align}
The lower bound is approached in the limit of an infinitely strong driving pulse, corresponding to \(s_\pi\rightarrow0\). In dimensional units, this implies that the minimum operational time approaches the total thermalization time, \[T^*_{\rm on-off}\rightarrow\tau_h+\tau_c.\]
\begin{figure}[t]
\centering
\includegraphics[width=.8\linewidth]{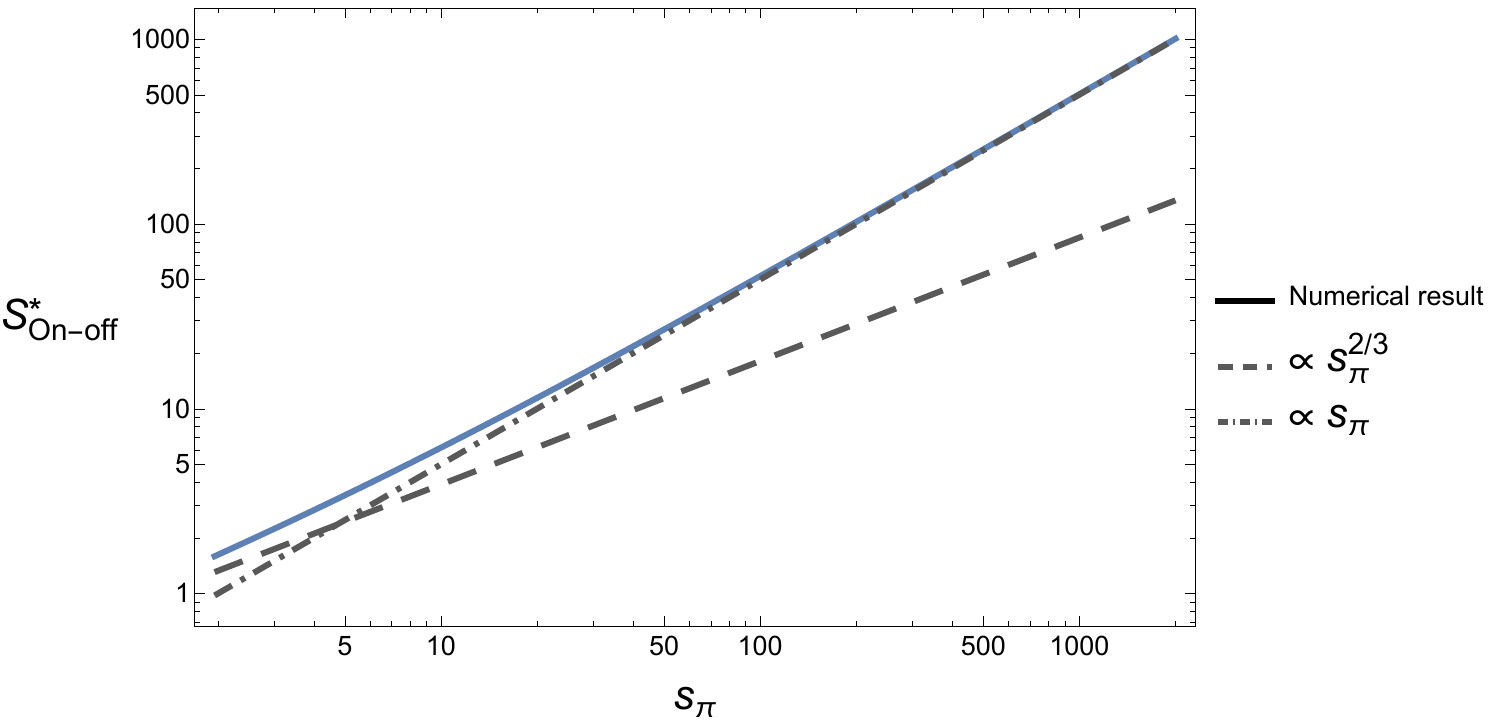}
\caption{
Log-log plot of the optimized dimensionless operational time \(S^*_{\rm on\text{-}off}\) as a function of the dimensionless modulation parameter \(s_\pi\), for $\kappa=1$. The solid curve corresponds to the numerical optimization, while the dashed and dot-dashed curves represent the asymptotic scalings \(S^*_{\rm on\text{-}off}-1\propto s_{\pi}^{2/3}\) and \(S^*_{\rm on\text{-}off}\propto s_{\pi}\), respectively. The logarithmic scales emphasize the crossover between the two asymptotic power-law regimes.}
\label{fig:asymptotic-onoff}
\end{figure}

Figure~\ref{fig:asymptotic-onoff} confirms that the optimized operating time exhibits the same asymptotic behavior as the four-stroke engine. In particular, the log-log representation reveals two distinct power-law regimes,
\begin{align}
S^*_{\rm on\text{-}off}-1\propto s_{\pi}^{2/3},
\qquad
s_{\pi}\ll1,
\end{align}
and
\begin{align}
S^*_{\rm on\text{-}off}\propto s_{\pi},
\qquad
s_{\pi}\gg1.
\end{align}
Although the on-off and four-stroke engines share a similar asymptotic behavior, the optimized on-off characteristic time has a smaller numerical prefactor than its four-stroke counterpart. Since $s_\pi$ and $s_{\rm LT}$ represent different physical control parameters, however, this difference should not be interpreted as an absolute performance advantage under identical experimental resources.

\section{Characteristic times}
For each engine, we can associate a characteristic timescale that determines its speed. For completeness, we summarize here the characteristic timescales and operational times derived for the four models.
\newline
    
Characteristic timescales:
\begin{align}
t_\pi&=\frac{\pi}{\Omega} \quad \text{(transverse coupling timescale)}\\
t_{\rm LT} &= t_{\rm LT, 1} + t_{\rm LT, 2} \quad \text{(level transformation timescale)} \\
%\tau_g &= \frac{2 \pi}{g} \quad \text{(harmonic modulation timescale)} \\
%\tau_{\rm F} &=\frac{2\pi}{\omega_{\rm F}}= \frac{4 \pi}{\omega_h - \omega_c} \quad \text{(internal dynamics timescale)} \\
\tau_{h,c} &= \frac{1}{\gamma_{h,c}^+ (\omega_{h,c}) + \gamma_{h,c}^{-}(\omega_{h,c})} \quad \text{(thermalization timescale)}
\end{align}
Operational timescales:
\begin{align}
T^*_{\rm 4-stroke} & = \min_{t_h,t_c} \left[\frac{(t_h+t_c+t_{\rm LT})(1 - e^{-({ t_h/\tau_h} +  t_c/\tau_c)})}{\left(1 - e^{ -t_h/\tau_h} \right) \left( 1 - e^{ -t_c/\tau_c} \right) }\right] \\
T^*_{\rm Floquet} &= \left(\tau_h + \tau_c\right) J^{-2}_1\left(\frac{\Omega}{\omega_{\rm F}}\right) \\
T^*_{\rm always-on} &= (\tau_h + \tau_c)\left(1+\frac{t_{\pi}^2}{\pi^2 \tau_h \tau_c} \right) \\
T^*_{\rm on-off} &= \min_{t_{hc}} \left[\frac{(t_\pi+t_{hc})(1 - e^{-t_{hc}\left(1/\tau_h +1/\tau_c \right)})}{\left(1 - e^{ -t_{hc} /\tau_h} \right) \left(1 - e^{ - t_{hc}/\tau_c} \right)} \right] 
\end{align}

Rescaled characteristic times:
% \begin{align}
%     T_{\rm 4-stroke}/\tau_h &= \frac{1 + \kappa}{2} \left(1+ \frac{s_{\rm LT}^2}{4 \pi^2 \kappa } \right) \\
%     T_{\rm Floquet}/\tau_h &= \min_x \left[\Big(2 s_{\rm LT}+x\Big)\frac{\left(1 - e^{ -2x(1+ \frac{1}{\kappa})} \right)}{\left(1 - e^{ -\frac{2x}{\kappa}} \right)\left(1 - e^{ -2x} \right)}\right] \\
%     T_{\rm Always-on}/\tau_h & = \min_{s_h,s_c} \left[\Big(2 s_F + s_h + \frac{s_c}{\kappa}\Big)\frac{\left(1 - e^{ -2(\frac{s_c}{\kappa}+s_h)} \right)}{\left(1 - e^{ -\frac{2s_c}{\kappa}} \right)\left(1 - e^{ -2s_h} \right)}\right] \\
%     T_{\rm On-off}/\tau_h &= \frac{1}{2}\left(1 + \kappa\right) J^{-2}_1\left(2s_\pi\right)
% \end{align}
% where $\kappa = \tau_c/\tau_h$, $s_{\rm LT} = \tau_\Omega / \tau_h$, $s_F = \tau_k/\tau_h$, $s_\pi = \tau_{\rm F}/\tau_g$
%Second version:
% \begin{align}
%     T_{\rm 4-stroke}/\tau_{\rm avg} &=2\left(1+ \left(\frac{\kappa s_{\rm LT}}{4 \pi}\right)^2\right) \ \\
%     T_{\rm Floquet}/\tau_{\rm avg} &= \min_x \left[\Big(2 s_{\rm LT}+x\Big)\frac{\left(1 - e^{ -2x \kappa^2} \right)}{\left(1 - e^{ -x \kappa^2(1 -\xi} \right)\left(1 - e^{ -x \kappa^2(1 +\xi} \right)}\right] \\
%     T_{\rm Always-on}/\tau_{\rm avg} & = \min_{s_h,s_c} \left[\Big(2 s_F + s_h + s_c + \xi(s_c-s_h)\Big)\frac{\left(1 - e^{ -(s_c + s_h)} \right)}{\left(1 - e^{ -s_c} \right)\left(1 - e^{ -s_h} \right)}\right] \\
%     T_{\rm On-off}/\tau_{\rm avg} &= 2 J^{-2}_1\left(s_\pi\right)
% \end{align}
\begin{align}
S^*_{\rm 4-stroke}(s_{\rm LT})&=T^*_{\rm 4-stroke}/(\tau_h + \tau_c)  = \min_{s_h,s_c} \left[\frac{\kappa}{2}\frac{(s_{\rm LT}+s_h+s_c)
\left(1-e^{-(\sigma s_h+s_c/\sigma)}\right)}{\left(1-e^{-\sigma s_h}\right)\left(1-e^{ s_c/\sigma}\right)}\right] \\
S^*_{\rm Floquet}(s_{\rm F})&=T^*_{\rm Floquet}/(\tau_h + \tau_c) = J^{-2}_1\left(s_F^{-1}\right) \\
S^*_{\rm always-on}(s_\pi)&=T^*_{\rm always-on}/(\tau_h + \tau_c) =1+\frac{s_\pi^2}{\pi^2}  \\
S^*_{\rm on-off}(s_\pi)&=T^*_{\rm on-off}/(\tau_h + \tau_c) = \min_{s_{hc}} \left[\frac{\kappa}{2}
\frac{(s_\pi+s_{hc}) \left(1-e^{-2s_{hc}/\kappa}\right)} {\left(1-e^{-\sigma s_{hc}}\right) \left(1-e^{-s_{hc}/\sigma}\right)}\right] 
\end{align}
where 
\begin{equation}
\kappa=\frac{2\sqrt{\tau_h \tau_c}}{\tau_h + \tau_c}, \quad \sigma=\sqrt{\frac{\tau_c}{\tau_h}},
\end{equation}
and
\begin{equation}
s_{\rm LT}=\frac{t_{\rm LT}}{\sqrt{\tau_h\tau_c}},
\quad
s_h=\frac{t_h}{\sqrt{\tau_h\tau_c}},
\quad
s_c=\frac{t_c}{\sqrt{\tau_h\tau_c}}
\quad
s_F=\frac{\omega_{\rm F}}{\Omega}
\quad
s_\pi=\frac{t_\pi}{\sqrt{\tau_h\tau_c}}
\quad
s_{hc}=\frac{t_{hc}}{\sqrt{\tau_h\tau_c}}.
\end{equation} 

\end{document}